\documentclass[acmsmall]{acmart}
\AtBeginDocument{%
  \providecommand\BibTeX{{%
    \normalfont B\kern-0.5em{\scshape i\kern-0.25em b}\kern-0.8em\TeX}}}
\setcopyright{acmlicensed}
\copyrightyear{2024}
\acmYear{2024}
\acmDOI{XXXXXXX.XXXXXXX}
\acmJournal{JACM}
\acmVolume{1}
\acmNumber{1}
\acmArticle{1}
\acmMonth{6}
\usepackage{orcidlink}
\usepackage{hyperref}
\usepackage{footmisc}
\usepackage[T1]{fontenc}
\usepackage{epstopdf}
\usepackage{graphicx}
\usepackage{placeins}
\usepackage{float}
\usepackage{nomencl}
\usepackage{color,soul}
\usepackage{multirow}
\usepackage{makecell}
\hypersetup{pdfauthor=author}
\usepackage{longtable}
\usepackage{hhline}
\usepackage{pifont}
\usepackage{graphics}
\usepackage{soul}
\usepackage{tabularx}
\usepackage[inline,shortlabels]{enumitem}
\usepackage{geometry}
\usepackage{xurl}
\usepackage{xspace}
\usepackage[binary-units,per-mode=symbol,exponent-product=\cdot,list-units=single,list-final-separator={,}]{siunitx}
\DeclareSIUnit[number-unit-product = ]\pixel{p}
\usepackage{chemformula}
\usepackage{gensymb}
\usepackage{amsmath}
\usepackage[inline]{enumitem}
\usepackage{booktabs}
\usepackage{comment}
\usepackage{footnote}
\usepackage{tablefootnote}
\usepackage{graphicx}
\usepackage{caption}
\usepackage{wrapfig, lipsum}
\usepackage{pifont}    
\expandafter\newcommand\csname r@tocindent4\endcsname{4in}
\newcommand{\etal}{et al.,\xspace}
\newcommand{\ie}{{i.e.}, }
\newcommand{\eg}{{e.g.}, }

\DeclareSIUnit{\pixelperwatt}{Pixel/W}
\newcommand{\HAS}{\emph{HTTP Adaptive Streaming }}
\newcommand{\HLS}{\emph{HTTP Live Streaming}}
\newcommand{\DASH}{\emph{Dynamic Adaptive Streaming over HTTP}}

\usepackage{tikz,xcolor,hyperref}
\definecolor{lime}{HTML}{A6CE39}
\DeclareRobustCommand{\orcidicon}{%
	\begin{tikzpicture}
	\draw[lime, fill=lime] (0,0) 
	circle [radius=0.16] 
	node[white] {{\fontfamily{qag}\selectfont \tiny ID}};
	\draw[white, fill=white] (-0.0625,0.095) 
	circle [radius=0.007];
	\end{tikzpicture}
	\hspace{-2mm}
}
\foreach \x in {A, ..., Z}{%
	\expandafter\xdef\csname orcid\x\endcsname{\noexpand\href{https://orcid.org/\csname orcidauthor\x\endcsname}{\noexpand\orcidicon}}
 }

\usepackage{ragged2e} 
\DeclareSIUnit[number-unit-product = ]\pixel{P}
\begin{document}
\title{Towards AI-Assisted Sustainable  Adaptive Video Streaming Systems: Tutorial and Survey}

\author{Reza Farahani}
\orcid{0000-0002-2376-5802}

\author{Zoha Azimi}
\orcid{0009-0007-5405-0643}

\author{Christian Timmerer}
\orcid{0000-0002-0031-5243}

\author{Radu Prodan}
\orcid{0000-0002-8247-5426}

\affiliation{
  \institution{Institute of Information Technology (ITEC), Alpen-Adria-Universität Klagenfurt}
  \city{Klagenfurt}
  \country{Austria}}
\email{firstname.surname@aau.at}
\renewcommand{\shortauthors}{Reza Farahani et al.}
\begin{abstract}
\vspace{-.7em}
\noindent\rule{\textwidth}{0.4pt}
Improvements in networking technologies and the steadily increasing numbers of users, as well as the shift from traditional broadcasting to streaming content over the Internet, have made video applications (e.g., live and Video-on-Demand (VoD)) predominant sources of traffic. Recent advances in Artificial Intelligence (AI) and its widespread application in various academic and industrial fields have focused on designing and implementing a variety of video compression and content delivery techniques to improve user Quality of Experience (QoE). However, providing high QoE services results in more energy consumption and carbon footprint across the service delivery path, extending from the end user's device through the network and service infrastructure (e.g., cloud providers). Despite the importance of energy efficiency in video streaming, there is a lack of comprehensive surveys covering state-of-the-art AI techniques and their applications throughout the video streaming lifecycle. 
Existing surveys typically focus on specific parts, such as video encoding, delivery networks, playback, or quality assessment, without providing a holistic view of the entire lifecycle and its impact on energy consumption and QoE.
Motivated by this research gap, this survey provides a comprehensive overview of the video streaming lifecycle, content delivery, energy and Video Quality Assessment (VQA) metrics and models, and AI techniques employed in video streaming. In addition, it conducts an in-depth state-of-the-art analysis focused on AI-driven approaches to enhance the energy efficiency of end-to-end aspects of video streaming systems (i.e., encoding, delivery network, playback, and VQA approaches). It finally discusses prospective research directions for developing AI-assisted energy-aware video streaming systems.
\end{abstract}


\maketitle
\noindent\rule{\textwidth}{0.4pt}
\section{Introduction}
\label{sec:intro}
In recent years, the utilization of the Internet has undergone a significant transformation, with video content now comprising approximately \qty{70}{\percent} of total Internet traffic. The Ericsson Mobility Report~\cite{ericsson} indicates that over \qty{80}{\percent} of all mobile data traffic will comprise video by \num{2028}, exceeding the current figure of \qty{71}{\percent}. There has been a notable increase in the demand for video streaming applications among all video traffic. For example, live video streaming accounts for more than \qty{17}{\percent} of all traffic in \num{2023}~\cite{Sandvine_report}. This growing demand for streaming applications, particularly after the COVID-19 pandemic, has intensified society's dependence on Internet services and contributed significantly to a \qty{20}{\percent} increase in the use of Over-the-Top (OTT) providers (e.g., YouTube, NetFlix) in \num{2020}~\cite{conviva}. The rise in demand, coupled with the support for higher quality and resolution content on modern smart devices across diverse domains such as entertainment, broadcasting, healthcare, and education, has motivated OTTs and Internet Service Providers (ISP) to improve further the user Quality of Experience (QoE) and network Quality of Service (QoS). These efforts can increase the energy consumption in the video streaming lifecycle, encompassing content capturing to rendering, consequently emitting a substantial volume of \ch{CO2} and other Greenhouse Gases (GHG). For this purpose, many initiatives, such as the Dimpact~\cite{dimapct} project, involve cooperative efforts of media, entertainment, and technology companies to diminish the environmental footprint and enhance the sustainability of the digital media industry.
\begin{table*}[!t]
\caption{Summary of related surveys.}
\label{tab:survey}
\resizebox{\textwidth}{!}{
\centering
\renewcommand{\arraystretch}{1.2}
    \begin{tabular}{|c|c|c|c|c|c|c|c|c|}
    \toprule
    \multirow{2}{*}{\emph{Survey}} & \multirow{2}{*}{\emph{Year}} &  \multirow{2}{*}{\emph{Scope}} & \multicolumn{4}{c|}{\emph{Video streaming components}} &  \multirow{2}{*}{\emph{AI}} & \multirow{2}{*}{\emph{Energy}}   \\ 
    \cline{4-7} 
    &&& \emph{Encoding} & \emph{Delivery} & \emph{Playback} & \emph{VQA} &&  \\
    \midrule
    \cite{ma2019image} & 2019 & Video compression for intra- or inter-frame prediction. & \ding{51} & \ding{53} & \ding{53} & \ding{53} & \ding{51} & \ding{53}\\
    \hline    
    
    \multirow{1}{*}{\cite{zhang2020machine}} & \multirow{1}{*}{2020} & Video compression for intra- or inter-frame prediction; video quality assessment.  & \multirow{1}{*}{\ding{51}} & \multirow{1}{*}{\ding{53}} & \multirow{1}{*}{\ding{53}} & \multirow{1}{*}{\ding{51}} & \multirow{1}{*}{\ding{51}} & \multirow{1}{*}{\ding{53}}  \\
    
   \hline
   
    \cite{shuja2021applying, makhkamov2023energy} & 2021, 2023 & Content caching and prefetching policies. & \ding{53} & \ding{51} & \ding{53} & \ding{53} & \ding{51} & \ding{53} \\   
    \hline 

    \multirow{2}{*}{\cite{saeik2021task, khan2022survey, feng2022computation}} & \multirow{2}{*}{2021, 2022} & Video compression and transcoding; caching, offloading, and distribution & \multirow{2}{*}{\ding{53}} & \multirow{2}{*}{\ding{51}} & \multirow{2}{*}{\ding{53}} & \multirow{2}{*}{\ding{53}} & \multirow{2}{*}{\ding{51}} & \multirow{2}{*}{\ding{51}} \\
    &  & between MEC and clouds. &  & & & &  &  \\
    \hline
    
    \cite{kumar2022machine} & 2022 & Scheduling and resource allocation techniques for video streaming. & \ding{53} & \ding{51} & \ding{53} & \ding{53}& \ding{51} & \ding{51}\\
    \hline 
    
    \cite{sani2017adaptive} & 2017 & ABR algorithm components and their interactions for HAS-based applications. & \ding{53}  & \ding{53} & \ding{51} & \ding{53} & \ding{51} & \ding{51} \\
    \hline

    \multirow{2}{*}{\cite{barakabitze2019qoe}} & \multirow{2}{*}{2019} &QoE modeling, measurement, and control in
    Software-Defined   & \multirow{2}{*}{\ding{53}}  & \multirow{2}{*}{\ding{51}} & \multirow{2}{*}{\ding{53}} & \multirow{2}{*}{\ding{53}} & \multirow{2}{*}{\ding{51}} & \multirow{2}{*}{\ding{51}} \\
     &  & Networking (SDN) using Network Function Virtualization (NFV).&  &  &  &  & &  \\
     \hline 

    \cite{borges2024systematic} & 2024 & Heuristic and AI-based methods for accelerating video transcoding process. &  \ding{51} & \ding{53} & \ding{53} & \ding{53} & \ding{51} & \ding{53} \\

    \hline 

    \multirow{2}{*}{\cite{afzal2024survey}} & \multirow{2}{*}{2024} & Energy consumption and environmental impact, as well as &  \multirow{2}{*}{\ding{51}} & \multirow{2}{*}{\ding{53}} & \multirow{2}{*}{\ding{51}} & \multirow{2}{*}{\ding{53}} &   \multirow{2}{*}{\ding{53}} & \multirow{2}{*}{\ding{51}} \\
    & & tools and datasets for encoding, decoding, and displaying video streaming. &  &  &  &  & & \\
    
    \toprule
    
    \multirow{2}{*}{This survey} & \multirow{2}{*}{2024} & Tutorial and survey of AI-based energy-efficient video streaming lifecycle &  \multirow{2}{*}{\ding{51}} &  \multirow{2}{*}{\ding{51}}  &  \multirow{2}{*}{\ding{51}} & \multirow{2}{*}{\ding{51}} & \multirow{2}{*}{\ding{51}}  & \multirow{2}{*}{\ding{51}} \\
    &  & i.e., encoding, decoding, delivery, playback, adaptive algorithms, quality assessment. &  & & & &  &  \\
    \bottomrule 
 \end{tabular}}
\end{table*}

Given the dynamic characteristics inherent in video streaming applications, reducing energy consumption and \ch{CO2} emissions while meeting users' QoE and latency expectations poses a considerable challenge. This challenge becomes more complex by quantifying and analyzing the energy consumption of video streaming applications, given the extensive data, diverse sequences, devices, and networks involved. Therefore, recent academic and industrial efforts like~\cite{xia2020qoe,xin2022ai,palit2023improving,Eric-AI} employ Artificial Intelligence (AI) to address these challenges and use its capabilities to manage large datasets, process structured and unstructured data, extract complex patterns, and provide predictions and insights.
Therefore, accurately identifying the appropriate stage in the video streaming lifecycle for applying AI techniques and a comprehensive understanding of the influential factors is critical.

Recent surveys on AI techniques aimed at improving the energy consumption of video streaming applications predominantly focus on specific stages of the streaming lifecycle, such as encoding, delivery, playback, or video assessment. As summarized in Table~\ref{tab:survey}, works like~\cite{ma2019image, zhang2020machine} focused on the video encoding part, discussing AI-based video compression techniques. In contrast, works like~\cite{shuja2021applying, makhkamov2023energy} studied AI-based in-network techniques such as content caching and popularity prediction. Other works such as~\cite{saeik2021task,kumar2022machine, khan2022survey, feng2022computation} surveyed AI-based solutions to investigate resource allocation and scheduling within the delivery network segment. In addition, Sani~\etal~\cite{sani2017adaptive} reviewed the literature on different Adaptive Bitrate Streaming (ABR) algorithms by investigating various techniques for monitoring and measuring resources such as network throughput, scheduling segment requests to servers, and adapting the video quality to match resources. 

Barakabitze~\etal~\cite{barakabitze2019qoe} presented a review of state-of-the-art QoE management solutions for multimedia streaming. Their work integrates modeling, monitoring, and optimization within emerging technologies and architectures, including cloud and edge computing, and extends to new domains such as immersive multimedia and video gaming. Borges~\etal~\cite{borges2024systematic} conducted a survey and categorized heuristic and AI solutions to accelerate the video transcoding process. Afzal~\etal~\cite{afzal2024survey} investigated the energy consumption and environmental impact of video streaming applications, focusing on the encoding, decoding, and display components, excluding AI-based solutions. They also discussed different tools and datasets commonly used in the literature for measuring and analyzing energy consumption in video streaming applications. However, existing surveys need a thorough literature review and taxonomy in the entire video streaming lifecycle to bridge the gap in exploring AI's contributions to optimizing energy consumption and QoE. The primary contributions of this survey are three-fold:
\begin{figure}[!t]
    \centering
    \includegraphics[width=\textwidth]{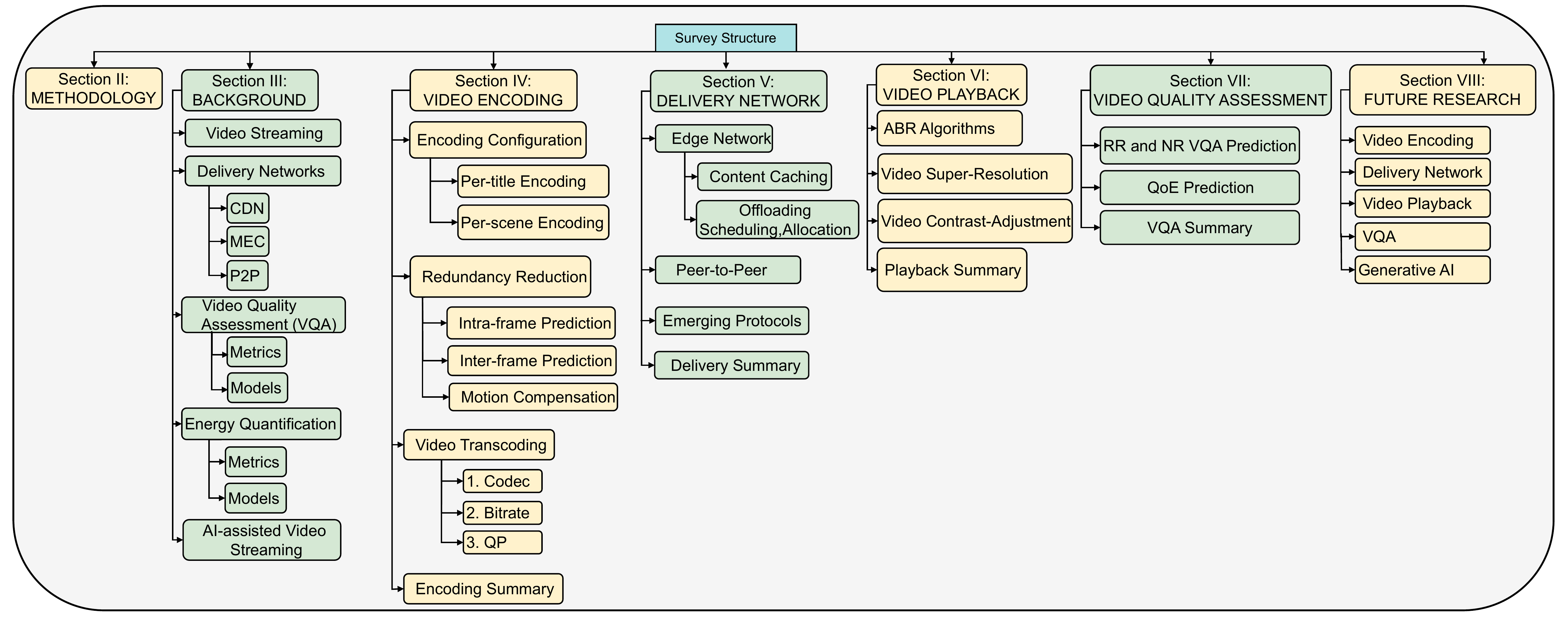}
    \caption{Survey structure.}
    \label{fig:structure}
\end{figure}
\begin{enumerate}[leftmargin=*]
    \item\textit{Tutorial} on video streaming lifecycle, including encoding, delivery, and playback and exploring metrics and models relevant to video QoE, energy, and AI techniques employed. 
    \item\textit{Taxonomy and survey} of \num{59} state-of-the-art AI-based energy-aware video streaming works, covering the entire streaming lifecycle and exploring the influence of predicting video quality assessment (VQA) metrics on minimizing energy consumption.
    \item\textit{Future directions} in designing AI-enabled energy-efficient video streaming systems based on the specific lifecycle parts identified in the survey.
\end{enumerate}

Fig.~\ref{fig:structure} shows the organization of this survey. Section~\ref{sec:methodology} describes the systematic methodology of the survey, followed by a background on video streaming, VQA and energy metrics and models, and AI-assisted video streaming systems in Section~\ref{sec:background}. Sections~\ref{subsec:encode}, \ref{subsec:delivery}, \ref{subsec:player}, \ref{subsec:vqa} provide a taxonomy and survey the state-of-the-art works in four categories: encoding, delivery network, playback, and VQA. Section~\ref{sec:challenge} outlines future research directions of AI-driven energy-efficient video streaming systems before concluding the article in Section~\ref{sec:concl}.

\section{Methodology}
\label{sec:methodology}
\begin{figure}[!t]
    \centering
    \includegraphics[width=1\linewidth]{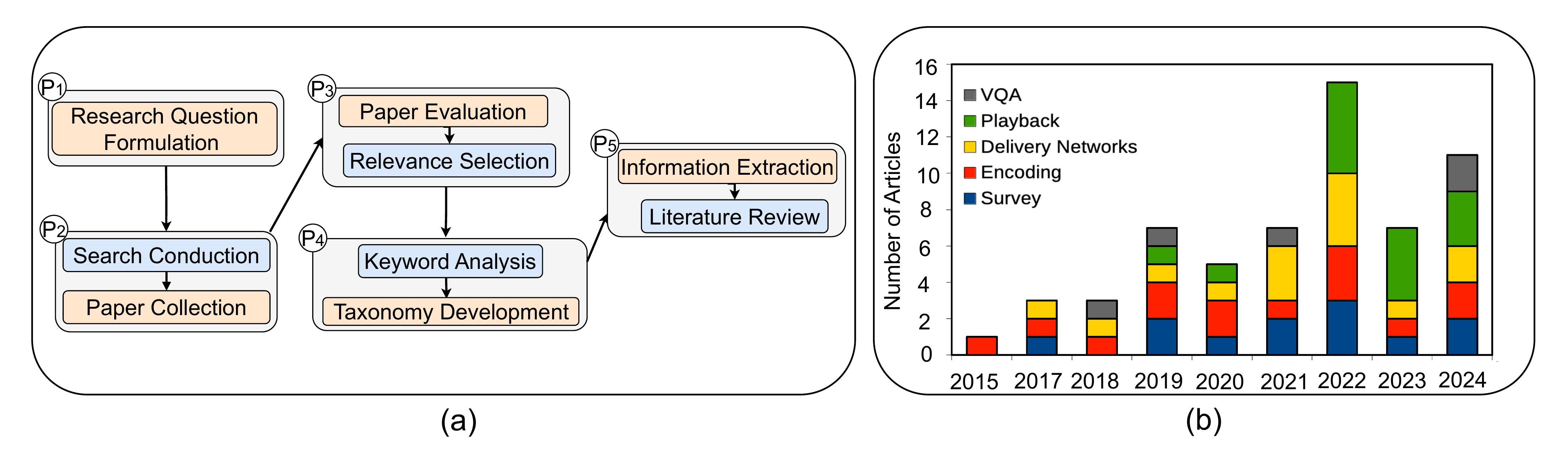}
    \caption{(a) Systematic literature review methodology, and (b) distribution of surveyed articles based on video streaming lifecycle segments across different years.}
    \label{fig:method}
\end{figure}

This section describes the systematic review methodology used in this survey, inspired by~\cite{keele2007guidelines} comprising five \emph{phases (P)} shown in Fig.~\ref{fig:method} (a).

\begin{enumerate}[leftmargin=*,label={\emph P\arabic*:}]
    \item{\emph{Research question formulation}.} We formulated five \emph{research questions (Q)} by delineating primary areas of the video streaming lifecycle. 
    \begin{enumerate}[label={\emph Q\arabic*)}]
     \item{\emph{Video encoding:}} 
        How can AI actively contribute to developing content-aware provisioning, considering both energy and QoE?
     \item{\emph{Video delivery:}} 
       How can AI techniques optimize content delivery services through decision-making policies, e.g., offloading and scheduling to balance network QoS and user QoE while enhancing energy efficiency and reducing economic costs?
     \item{\emph{ABR algorithm:}} 
        How can AI enhance ABR algorithms by integrating content-aware information to improve bitrate selection and energy efficiency?
     \item{\emph{Video playback:}}
        How can client-based AI technologies, such as super-resolution and denoising, optimize video playback while balancing quality enhancement with energy efficiency?
     \item{\emph{VQA:}}
        How can AI-based techniques reduce dependency on reference frames in computational VQA processes to optimize energy consumption?
    \end{enumerate}
\item{\emph{Search and paper collection}.} We conducted comprehensive searches in the scholarly databases and venues listed in Table~\ref{tab:method} considering the five research questions and collected \num{850} references.
\item{\emph{Relevant paper selection.}}
In many cases, determining the relevant articles based on their titles proved challenging. Consequently, we implemented additional filters by scrutinizing the abstracts and meticulously reviewing papers that met the following criteria: (1) published after 2015, indicative of a significant trend in the research landscape; (2) sourced from 
journals with high impact factors or conferences ranked B or higher according to the \emph{CORE} ranking pertinent to the field (Table~\ref{tab:method}); and (3) authored by individuals affiliated with universities ranked in the top \num{500} of the \emph{QS World University Rankings} or companies listed in the \emph{Fortune 500}.
\item{\emph{Keyword analysis and taxonomy development.}}
We used \num{100} keywords from the video, energy, and AI domains to conduct a detailed analysis and formulate a taxonomy for AI-assisted sustainable systems, classified into four categories: encoding, delivery, playback, and VQA. 
\item{\emph{Information extraction and literature review.}}
We extracted information from the selected works to address the five research questions outlined in the first phase. We contacted the authors to gain more in-depth information about their proposed methodologies and experimental designs when required. From the initially selected \num{850} articles, we thoroughly surveyed \num{210} articles, referencing \num{59} articles that directly or indirectly use AI-based energy-aware solutions for the video streaming lifecycle (Fig.~\ref{fig:method} (b)). 
\end{enumerate}
\begin{table*}[!t]
\centering
\caption{Search strategy and sources in the systematic literature review.}
\label{tab:method}
\resizebox{\textwidth}{!}{
\fontsize{7pt}{7pt}\selectfont
\renewcommand{\arraystretch}{1.2}
\begin{tabular}{|c|cl|}
\toprule
\textit{Year}   & \multicolumn{2}{c|}{\numrange{2018}{2024}}                      \\ \hline
\multirow{3}{*}{\textit{Keywords}}  
& \multicolumn{1}{c|}{Video} & \begin{tabular}[l]{@{}l@{}}
\textbf{General}: Video, video streaming, VoD, live, HTTP adaptive streaming, MPEG DASH, Apple HLS, AVC, HEVC, \\VVC, AV1, VP9. \\ \hline  \textbf{Encoding}: Entropy encoding, transform coding, 
intra-, inter-frame, motion estimation/compensation, CU, PU, \\ CTU, 
partitioning, block partitioning, per-title encoding, per-scene encoding, transcoding, transrating. \\ 
\hline \textbf{Delivery}: Video transmission, network-assisted streaming, CDN, edge computing, task scheduling, MEC, CDN,\\ computing continuum, WebRTC cloud computing, content delivery, resource allocation, P2P, D2D, TCP, QUIC,\\  MPTCP,  content caching. \\ 
\hline \textbf{Playback}: video decoding, ABR, quality enhancement techniques, super-resolution, frame interpolation, \\denoising,  display.\\ 
\hline  \textbf{VQA}: VMAF, SSIM, PSNR, QoE, latency, P1203, video quality assessment, stalling events, rebuffering, \\quality switch.
\end{tabular}   \\ \cline{2-3}
& \multicolumn{1}{c|}{Sustainability} & \begin{tabular}[l]{@{}l@{}}
Energy consumption, power consumption, battery, battery life, \ch{CO2}, carbon footprint, green, sustainable, \\energy efficiency. 
\end{tabular}   \\ \cline{2-3}
& \multicolumn{1}{c|}{AI}  &  \begin{tabular}[l]{@{}l@{}}
Machine learning, supervised learning, unsupervised learning, decision tree, random forest, SVM, LLM, \\prediction, classification, deep learning, CNN, LSTM, reinforcement learning, Q-learning, online learning \\deep reinforcement learning, generative AI, explainable AI.           \end{tabular}       \\
\midrule
\multirow{2}{*}{\textit{Venues}}                 
& \multicolumn{1}{c|}{Encoding, Playback,}       &   
\textbf{IEEE:} TCSVT, TIP, Broadcasting, TMM, Access, DCC, ICME, MMSP, ICIP, VCIP, QoMEX. \\
& \multicolumn{1}{c|}{and VQA}       &  \textbf{ACM:} MM, MMM, TOMM. 
   \\ 
\cline{2-3}

& \multicolumn{1}{c|}{Delivery}       &   \begin{tabular}[l]{@{}l@{}}
\textbf{IEEE:} TNSM, TMC, Access, TPDS, ICC, ISM, LCN, GLOBECOM, INFOCOM, JSAC, TVT; \textbf{USENIX:} NSDI, ATC;  \\
\textbf{ACM:} SIGCOMM, CoNext, MMSys; \textbf{Elsevier:} Computer Networks, Computer Communications.\\
 
\end{tabular}   \\

\cline{2-3}

& \multicolumn{1}{c|}{\textit{Surveys}}  & \textbf{IEEE:} COMST, Access, TCSVT; \textbf{ACM:} CSUR, TOMM; \textbf{Elsevier:} Computer Networks, SUSCOM.        \\ 
\midrule

\multicolumn{2}{|c|}{\textit{Scholarly search engines}}  & \multicolumn{1}{c|}{Google Scholar, IEEE Xplore, ScienceDirect, ACM Digital Library, ResearchGate, arXiv.}        \\ 

\bottomrule
\end{tabular}}
\end{table*}
\section{Background}
\label{sec:background}
This section provides basic background knowledge on video streaming, delivery networks, VQA, and energy metrics and models. It also presents an overview of the AI-assisted techniques commonly employed for video streaming applications.
\subsection{Video Streaming}\label{sec:background:st}
Video streaming requires simultaneously downloading and playing back audio and video content, following the lifecycle in
Fig.~\ref{e2e}. In contrast to traditional video downloading, which retrieves the entire video sequence before playback, video streaming enables users to start watching almost instantly during video transmission. \HAS (HAS), e.g., based on Moving Picture Experts Group (MPEG) \DASH~(DASH) standardization~\cite{dash} or Apple \HLS~(HLS)~\cite{hls}, is nowadays the predominant video streaming technique~\cite{yin2015control,yaqoob2020survey}. HAS-based streaming encodes the video content in multiple quality (e.g., bitrate and/or resolution) levels, known as representations, and divides them into short and fixed-duration segments. A collection of these representations, known as the \emph{bitrate ladder}, is stored on the media server, such as the origin server or Content Delivery Network (CDN). A manifest file, such as the HLS playlist or DASH MPD, contains the representation information, including the addresses (i.e., HTTP Uniform Resource Locators (URLs)) of the segments stored on media servers. 
When the video player issues a streaming request, it loads this manifest file, analyzes it, and selects an appropriate segment representation. Determining proper representations depends on network throughput and player characteristics, such as buffer status and display resolution. Therefore, the client must continuously monitor these metrics and dynamically adjust the video quality by choosing the most appropriate representation for the next segment using an ABR algorithm to provide a smooth streaming session with a high network QoS and user QoE~\cite{herglotz2022modeling,hossfeld2023greener}.
\subsection{Delivery Networks}\label{sec:background:dli}
\begin{figure*}[!t]
\captionsetup{skip=10pt}
\centering
\includegraphics[width=\textwidth]{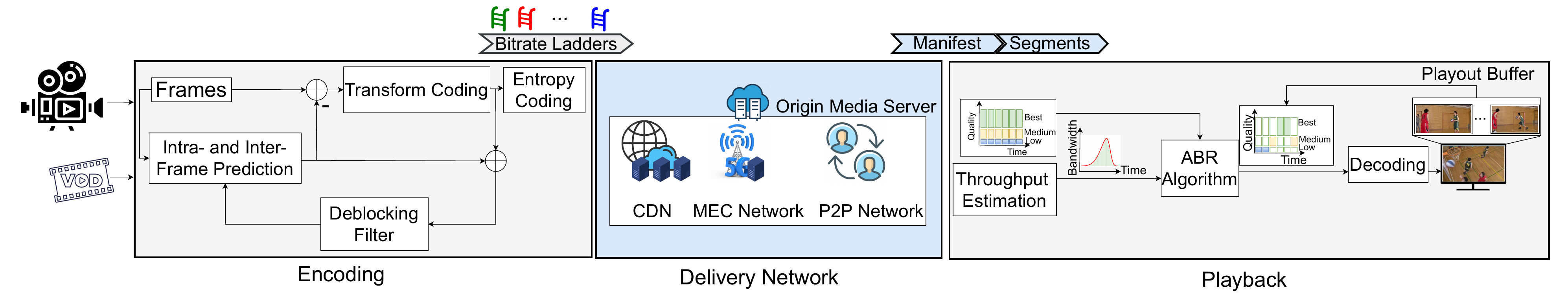}
\caption{General video streaming life cycle, including video encoding, content delivery, and playback phases.}
\label{e2e}
\end{figure*}

The delivery network includes an origin server with one or a hybrid combination of three networks, as shown in Fig.~\ref{e2e}.
\subsubsection{Content Delivery Network (CDN)} It is a distributed network of servers that work together to deliver content to users, including three main components~\cite{pathan2007taxonomy,nygren2010akamai}:
\begin{enumerate}
    \item \emph{CDN servers}, also known as surrogate, cache, or edge servers, placed in various geographical locations near end-users, cache and deliver content quickly and efficiently. A user request contacts the closest CDN server, and if the CDN server has the content in its cache, it fulfills the request faster than retrieving it from the origin server. Otherwise, the content is fetched from the origin server and served at rates slower than delivery from the CDN servers.
    \item \emph{Origin server} is responsible for storing and updating the original content. When a CDN server does not cache the requested content, the origin server serves the request and then feeds the CDN servers for future demands based on the content distributor policy.
    \item\emph{CDN content distributor} manages content distribution across the CDN and defines important policies for improving the network performance. For example, it monitors the CDN network to detect frequently accessed content, which it caches on CDN servers to reduce the load on the origin server and improve the overall content delivery performance.
\end{enumerate}
\subsubsection{Multi-access Edge Computing (MEC) network} As a complementary technology to cloud computing, provides storage and compute resources close to end-users at the network’s edge with lower data processing latencies, better user experience, and lower cost and bandwidth consumption~\cite{taleb2017multi}. MEC can offer several merits in various types of video streaming applications, including:
\begin{enumerate}
\item\emph{VoD streaming.} Applying techniques such as most popular content caching or content prefetching in advance at the edge enhances the scalability and reliability of VoD services, while reducing the load on cloud servers (e.g., origin server), enabling faster content delivery and boosting user QoE~\cite{aguilar2023space}.
\item\emph{Live streaming.} Caching, transcoding, and processing content closer to the user who generates and consumes the data reduces latency and improves live video streaming quality~\cite{farahani2022leader,da2024super}. 
\item\emph{Immersive video streaming.} Augmented Reality (AR) or Virtual Reality (VR) applications necessitate real-time processing and rendering of content to reduce latency and enhance performance. Thus, they demand advanced processing services and techniques beyond traditional streaming, such as real-time data analytics at the network edge closer to the users.
\end{enumerate}
\subsubsection{Peer-to-Peer (P2P) network} It is a decentralized delivery where each peer device communicates and shares content directly with others rather than relying on a centralized server. Each peer in a P2P network acts as both client and server that simultaneously downloads and uploads content~\cite{siano2019survey}. A typical P2P network includes three peer types:
\begin{enumerate}
\item\emph{Tracker server} records peers' logs and coordinates content exchange for efficient distribution and synchronization. When a peer requests content, it contacts the tracker server to fetch a list of others holding the content for direct download without relying on a central server. 
\item\emph{Seeders} are peers who completely downloaded and shared the content with others. Once a peer downloads the content from a seeder, it may become a seeder and continue sharing. Holding high resources, \eg upload/download bandwidth and battery, is another characteristic of seeders, which allows them to remain connected to the P2P network for extended periods; consequently, other peers fetch their desired content from these stable peers.
\item\emph{Leechers} are peers still downloading the content and have not yet fetched it completely. 
\end{enumerate}
\subsection{Video Quality Assessment (VQA)}\label{sec:background:vqa}
Efforts to reduce the energy consumption of video streaming applications must primarily preserve the users' QoE. Therefore, several standards and frameworks from the literature evaluate the quality of video sequences received by end users through QoE metrics and models.
\subsubsection{VQA metrics}\label{sec:VQAMetric}
They measure and assess the perceived quality of videos, facilitating comparisons between different techniques, and typically fall into two groups.
\begin{enumerate}
    \item{\emph{Subjective VQA metrics}} relate to human perception. \textit{Mean Opinion Score} (MOS)~\cite{mos} a popular metric that evaluates video QoE based on user opinions on a scale of \numrange{1}{5} (highest quality).
    \item{\emph{Objective VQA metrics}} use different algorithms to predict and quantify human perception based on parameters affecting the quality of a compressed video. 
\begin{itemize}
\item\textit{Peak Signal-to-Noise Ratio} (PSNR)~\cite{hore2010image} reports the mean squared error of differences in pixels of an original (signal) and compressed or distorted (noise) video. The PSNR is typically in the range of \qtyrange{30}{50}{\deci\bel} for \num{8}-bit data, and \qtyrange{60}{80}{\deci\bel} for \num{16}-bit data\cite{deshpande2018video}, where a higher value indicates a better quality. XPSNR~\cite{helmrich2020xpsnr} extends naive PSNR for ultra-high-resolution content. 
\item\textit{Structural Similarity} (SSIM)~\cite{wang2004image}
rather than using pixel information, computes the correlation of neighboring pixels and the match of the luminance and contrast of the compressed and original videos in the range \numrange{-1}{1} (where \num{1} shows perfect similarity)~\cite{lee2017comparison, luo2019vmaf}.
\item\textit{Video Multi-Method Assessment Fusion} (VMAF)~\cite{vmafgit} developed by Netflix  
extracts spatial, temporal, and human perceptual features from distorted and reference video frames. It uses such features as inputs to a Machine Learning (ML) model to predict the subjective quality that minimizes the difference to human judgments. VMAF ranges from \numrange{0}{100}, where a higher score indicates better quality~\cite{vmafblog}. 
\end{itemize}
\end{enumerate}

\subsubsection{VQA models}\label{sec:VQAModel}
VQA metrics do not consider some of the influential factors significantly affecting viewer satisfaction~\cite{barman2019qoe} such as:
\begin{itemize}
\item{\emph{Rebuffering/stalling}} interrupts or pauses video playback when the player runs out of buffered data and needs to load more to continue playback.
\item{\emph{Quality switches}} alterations in visual resolution or clarity of a video presentation.
\item{\emph{Startup delay}} represents the time lapse between initiating the playback request and the actual display of the video on the screen.
\end{itemize}
In recent years, several QoE models have been introduced to consider and incorporate these influential factors. The most well-known models are: 
\setcounter{paragraph}{0}
\begin{enumerate}
\item{\emph{ITU-T P.1203}} introduced by ITU~\cite{p1203} has three modules for estimating video and audio quality and audiovisual integration. This objective model quantifies QoE on a scale of \numrange{1}{5}, where a higher score indicates a higher QoE. The video quality estimation model has four operation modes with incremental access to the input information:
\begin{itemize}
\item \textit{Mode 0} accesses the codec, target bitrate, display resolution, framerate, and segment duration information;
\item \textit{Mode 1} considers frame sizes and frame type information in addition to input from mode 0;
\item \textit{Mode 2} receives up to \qty{2}{\percent} of bitstream information;
\item \textit{Mode 3} has access to all bitstream information on top of all the inputs of previous modes~\cite{robitza2018http}.
\end{itemize}
\item{\emph{Yin}~\etal~\cite{yin2014toward}} evaluates individual video segments using a weighted sum of video quality, quality switches, and total rebuffering time. Numerous studies, \cite{bentaleb2022bob, belda2020look} expanded its scope to include additional influential VQA factors, such as playback speed.
\end{enumerate}
\subsection{Energy Quantification}\label{sec:background:EE}
Efforts to quantify energy consumption and carbon emissions have gained significant momentum in recent years. Several works from the literature present metrics and models to measure and predict energy use and emissions across the video streaming lifecycle.
\subsubsection{Energy Metrics}\label{sec:energymetric}
The most common metrics used for measuring energy consumption are:
\begin{itemize}
\item \textit{Power consumption} measures the amount of electrical power consumed during video streaming in \unit{\kilo\watt}~\cite{yue2022eesrnet}.
\item \textit{Energy consumption} measured in \unit{\kilo\watt\hour} or \unit{\joule} is the total streaming energy used, usually by multiplying the power consumption with the video duration~\cite{li2021collaborative,bentaleb2023meta}.
\item \textit{Energy efficiency}
is the ratio between video quality (\eg resolution, bitrate) to the energy consumed, such as \unit[per-mode=symbol]{\bit\per\joule} or \si{\pixelperwatt}.
\item \textit{Battery life} assesses the duration for each device (\eg mobile or tablet) maintaining video streaming functionality, typically in hours, before recharging.
\item \textit{Carbon intensity} represents the average grams of \ch{CO2} equivalent emitted per \si{\kilo\watt\hour} of electricity generated with significant variation between countries and energy sources.
\item \textit{Carbon footprint} is the electricity consumption multiplied by the carbon intensity, reported as the total amount of GHG expressed in tonnes or \unit{\kilo\gram} of \ch{CO2} equivalent (\ch{CO2}e)~\cite{stephens2021carbon}.
\end{itemize}
\subsubsection{Energy models}\label{sec:energymodel}
Several recent energy consumption and carbon emissions models analyze and predict the impact of video streaming applications. Each model focuses on a specific stage of the video streaming lifecycle rather than adopting a comprehensive end-to-end approach.
\begin{enumerate}
    \item \emph{Encoding models} formulate or predict the energy requirements of compressing video sequences using various codecs and parameters. Sharrab~\etal~\cite{sharrab2013aggregate} introduced a linear model for encoding that specifically addresses the AVC codec utilized in live streaming systems incorporating encoding parameters, such as the number of reference frames, motion estimation range and algorithm, and Quantization Parameter (QP). Azimi~\etal~\cite{qomex2024} proposed an Extreme Gradient Boosting (XGBoost) model to predict the energy consumption of the encoding process by utilizing features such as video complexity, QP, resolution, framerate, and codec type.
    \item \emph{Delivery models} formulate or predict the energy required for preparing and transmitting video sequences across delivery networks to end users. Huang~\etal~\cite{huang2012close} proposed a linear model for LTE networks based on the uplink and downlink throughput. Mehrabi~\etal~\cite{mehrabi2019energy} modeled the energy consumption of an edge server for transcoding tasks in a time-slotted manner based on the data volume processed for all active clients. To this aim, they considered specific bitrate allocation to each client, completed transcoding tasks, a processing weight factor for each edge server, and the energy consumption per bit of data, accounting for interactions with other edge servers and the overall processing load.
    \item \emph{Playback models} formulate or predict the energy required to process, decompress, and playout content on user devices. Herglotz~\etal~\cite{herglotz2015estimating} introduced a linear regression model to predict energy consumption based on decoding processing time, enabling measurements independent of hardware and operating systems. Turkkan~\etal~\cite{turkkan2022greenabr} proposed an Artificial Neural Network (ANN) model to predict the power consumption of a video sequence during playback based on parameters such as bitrate, resolution, framerate, and file size. 
\end{enumerate}
\begin{figure}[!t]
    \centering
    \includegraphics[width=1\textwidth]{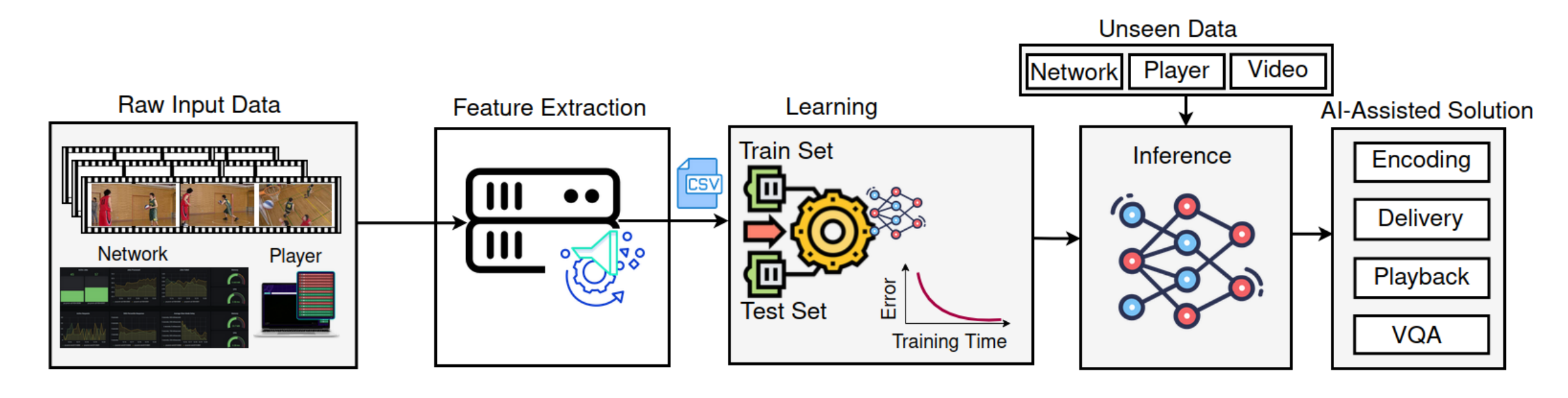}
    \caption{General AI-assisted video streaming workflow.}
    \label{fig:workflow}
\end{figure}
\subsection{AI-assisted Video Streaming}\label{sec:ai}
The typical pipeline of an AI-based video streaming system includes four stages, which can apply to one or multiple parts of the video streaming lifecycle (Fig.~\ref{fig:workflow}):
\begin{enumerate}[leftmargin=*]
    \item {\emph{Raw input data}} encompass details of the video sequence (\eg video complexity features), network (\eg throughput, load), CDN statistics (\eg cache occupancy, request popularity), and end-user devices (\eg playback state, buffer level, battery status)~\cite{farahani2023sarena,farahani2022ararat}. To address communication challenges among clients, the origin server, and the CDNs, the \textit{Consumer Technology Association} (CTA)~\cite{cta} in the \textit{Web Application Video Ecosystem} (WAVE) project has introduced the \textit{Common Media Client Data} (CMCD)~\cite{cmcd} and \textit{Common Media Server Data} (CMSD)~\cite{cmsd} specifications that standardize the streaming components to convey the information, ensuring consistent data interpretation and analysis.
    \item {\emph{Feature extraction}} involves identifying, capturing, and selecting relevant characteristics within the input data, including recognizing motion patterns, identifying recurring themes, and noting variations in the resolution of the video sequence. To facilitate this feature selection process, statistical models~\cite{khan2009content, seufert2019features} or machine learning algorithms~\cite{yu2021convolutional} are often applied, ensuring that the model prioritizes crucial elements to improve the desired output, e.g., video streaming quality and user satisfaction.
    \item {\emph{Learning}} involves training models to discern patterns within existing data and determine an appropriate decision-making action. This could entail \emph{Reinforcement Learning} (RL)~\cite{farahani2022hybrid,bentaleb2023meta}, classification using methods such as \textit{Support Vector Machines} (SVM)~\cite{zhang2015machine, zhang2017effective}, or prediction through techniques like \textit{Long Short-Term Memory} (LSTM)~\cite{menon2023transcoding, zheng2021hybrid} or \textit{Random Forest} (RF)~\cite{amirpour2023optimizing, menon2024energy}. The initial step divides the dataset into training and testing sets, with subsequent training of models on the designated training set. The model parameters are adjusted based on the features extracted in the previous stage, facilitating the understanding of the underlying patterns. To assess the model's efficacy on the test set, relevant evaluation metrics such as \textit{Mean Absolute Error} (MAE) or \textit{Mean Squared Error} (MSE) are defined~\cite{amirpour2023optimizing, menon2024energy}.
    \item {\emph{Inference}} utilizes the learned patterns and information to make decisions, predict, or classify unseen data associated with encoding, content delivery, playback, or VQA.
\end{enumerate}

\section{Video Encoding} 
\label{subsec:encode}
The encoding phase in video streaming compresses video sequences, facilitating efficient storage and seamless data transfer across the delivery network. This involves eliminating redundant data from raw input files or high-quality encoded video sequences and retaining only essential information. Various compression-decompression technologies, known as codecs, such as \textit{Advanced Video Coding} (AVC/H.264~\cite{h264, wiegand2003overview}), \textit{High Efficiency Video Coding} (HEVC/H.265~\cite{h265, sullivan2012overview}), \textit{AOMedia Video 1 (AV1)}~\cite{av12021}, and \textit{Versatile Video Coding} (VVC/H.266~\cite{h266, bross2021overview}), incorporate distinct compression modules (refer to Section~\ref{redundancy} for further information). Optimizing the encoding parameters by modules within these codecs reduces the data to encode, subsequently reducing the encoding time and computation complexity. This, in turn, contributes to energy efficiency in computation.
\subsection{Encoding Configuration}\label{encod-config}
Encoding configuration involves the choice and optimization of encoding parameters, in some implementations~\cite{wieckowski2021vvenc, x265} known as ``preset'', varies from \texttt{very slow} to \texttt{ultrafast}. Faster presets achieve faster encoding with less compression efficiency, while slower presets obtain higher compression efficiency at the cost of longer encoding time and more energy consumption~\cite{laude2019comprehensive}.
In addition, several other parameters influence compression efficiency, including:
\begin{itemize}
\item{\emph{Bitrate}} is the amount of data transmitted per unit of time, expressed in bits per second (bps). 
\item{\emph{Framerate}} is the number of individual frames displayed per second in a video, expressed in frames per second (fps).
\item{\emph{Resolution}} is the number of pixels in a video frame, expressed as the width and height.
\item{\emph{Quantization parameter (QP)}} balances video quality and compression efficiency by deciding which details to retain or discard during encoding process.
\end{itemize}

Conventional methods often rely on predetermined presets, independent of the video content and not tailored to the specific characteristics of video content, proving inefficient for video sequences with unique features~\cite{eusipco2024}. Video content featuring a high level of dynamism, characterized by rapid scene changes and fast-paced actions, differs significantly from static content. There are several works on customizing settings for each video title or scene, referred to as \textit{per-title} and \textit{per-scene} encoding, respectively. 
\subsubsection{Per-title encoding}\label{pertitle}
introduced by Netflix investigates the characteristics of video sequences for finding the ``best'' encoding parameters balancing quality and compression efficiency based on their complexity~\cite{de2016complexity}, encompassing scene changes, motion, and content variability.
For the content of average complexity with a \num{1080}p resolution, per-title encoding allocates \qty{20}{\percent} less bitrate compared to the fixed ladder method~\cite{netflixblog}. Although the impact of this method is significant in bitrate savings, it tends to be slower and more resource-intensive~\cite{mux_encoding}. Therefore, several research works are trying to improve the efficiency of per-title encoding, exploring AI-based approaches that leverage video feature extraction and optimized encoding parameter prediction capabilities.
Fig.~\ref{fig:pertitle} depicts an example workflow of AI-based per-title encoding that starts with live or VoD video source, including three key stages:
\begin{enumerate}[leftmargin=*]
\item{\emph{Content complexity analysis}} considers information like:
\begin{figure*}[!t]
    \centering
    \includegraphics[width=\textwidth]{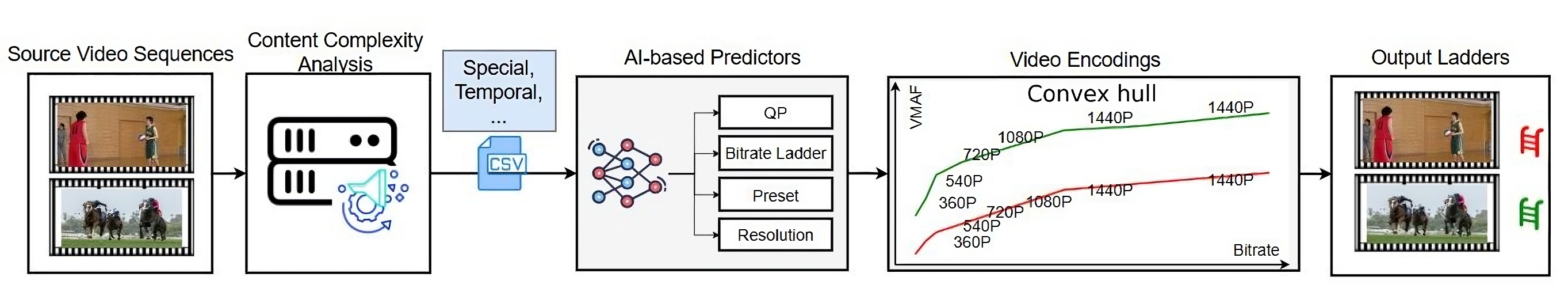}
    \caption{AI-based per-title encoding for two video sequences with distinct complexities.}
    \label{fig:pertitle}
\end{figure*}
\begin{enumerate*}
\item\textit{spatial information}, referring to the arrangement and distribution of pixels in video frames, including colors, shapes, textures, luminance, chrominance, and patterns;
\item\textit{temporal information}, relating to the variations over time in a video and capturing motion information and dynamics across frames.
\end{enumerate*}
\item{\emph{AI-based predictors}} uses ML and ANN-based models (e.g., Gaussian Process (GP)~\cite{katsenou2021efficient}, RF~\cite{menon2024energy}, and~\cite{falahati2024efficient}) to estimate the optimized encoding profile based on the complexity analysis.
\begin{itemize}
\item{\emph{QP prediction.}} Katsenou~\etal~\cite{katsenou2021efficient} proposed a content-driven method that extracts temporal and spatial features from the uncompressed content for each video sequence. It then applies a GP model to predict the QPs, leading to an optimized bitrate ladder that saves \qty{89.06}{\percent} of the required encodings compared to the exhaustive search method, leading to less processing time and energy consumption.
\item{\emph{Bitrate ladder prediction.}} Mueller~\etal~\cite{mueller2022context} proposed an encoding method using a combination of three ML models (i.e., XGBoost, \textit{Convolutional Neural Networks} (CNN), and fully connected neural networks to extract features from a video sequence based on its characteristics. They proposed a bitrate ladder that saves up to \qty{20}{\percent} bitrate and reduces the carbon footprint of video streaming compared to conventional per-title encoding.
\item{\emph{Preset prediction.}} Amirpour~\etal~\cite{amirpour2023optimizing} used an RF model to predict the optimized preset based on the brightness and temporal complexity of \num{500} video sequences for an HEVC encoder, showing \qty{70}{\percent} energy savings with a penalty of only \num{0.15} in the VMAF score.
\item{\emph{Resolution prediction.}} Menon~\etal~\cite{menon2024energy} introduced an RF-based prediction of the video encoding time and XPSNR, allowing dynamic resolution adjustments based on video segments' spatial and temporal characteristics. The approach achieved a bitrate reduction of \qty{12.03}{\percent} within a \qty{200}{\second} encoding time constraint with an impressive \qty{84.17}{\percent} decrease in encoding energy compared to the reference HLS bitrate ladder. Azimi~\etal~\cite{qomex2024} employed an XGBoost model to predict and balance encoding energy and VMAF scores using spatial and temporal complexity features, resolution, framerate, QP, and codec.
They used predictive models to identify optimal encoding parameters (i.e., resolution and QP) for each video sequence, balancing energy consumption and VMAF. 
The approach achieved an average energy reduction of \qty{46}{\percent} with only a four-unit decrease in VMAF.
\end{itemize}
\item{\emph{Video encoding}}
considers the content characteristics of each video by adjusting parameters obtained from the previous stage. After encoding, the \textit{convex hull}, which encapsulates the bitrate-resolution pairs with their corresponding VMAF scores, ensures that the selected combinations lie close to this convex shape. Fig.~\ref{fig:pertitle} shows two video sequences with distinct VMAF scores for identical bitrate and resolution values, underscoring the importance of different bitrate ladders for each video content based on its complexity.
\end{enumerate}
\subsubsection{Per-scene encoding}
\label{perscene}
It analyzes video sequences scene-by-scene and determines the best encoding parameters on their complexity and characteristics. Similar to per-title encoding, AI could be influential in analyzing the video sequence and extracting the features of each scene. For instance, \cite{seeliger2022green} used the Deepencode model~\cite{mueller2022context} to categorize the scenes in a video sequence based on their complexity and predict the best bitrate accordingly, which reduced the energy consumption of HD and UHD video by \qty{9}{\percent} and \qty{30}{\percent}, respectively. 

\subsection{Redundancy Reduction}
\label{redundancy}
Redundancy reduction minimizes the data necessary to represent video frames. Each video codec employs a set of modules. Since video sequences encompass both spatial and temporal features requiring compression, recent video coding standards incorporate hybrid {video coding techniques covering both spatial and temporal aspects~\cite{li2020explainable}, including:}
\begin{enumerate}
    \item\emph{Predictive coding} eliminates unnecessary video sequence information by addressing spatial and temporal redundancies utilizing intra- and inter-frame prediction methods. Higher prediction accuracy results in encoding fewer residuals, \ie differences between actual and predicted frames, leading to enhanced compression efficiency and energy consumption.
    \item\emph{Transform coding} {shifts residuals to a spectral domain. It then} quantizes the spectral coefficients to diminish spatial and perceptual redundancies. 
    \item\emph{Entropy coding} addresses statistical redundancies by encoding symbols with higher probabilities using fewer bits and symbols with lower probabilities with more bits.
\end{enumerate}
Many state-of-the-art codecs and works leverage advanced predictive compression techniques, including intra-frame and inter-frame coding and motion compensation.
\subsubsection{Intra-frame prediction}
\label{intraframe}
\begin{figure}[t]
    \centering
    \includegraphics[width=\textwidth]{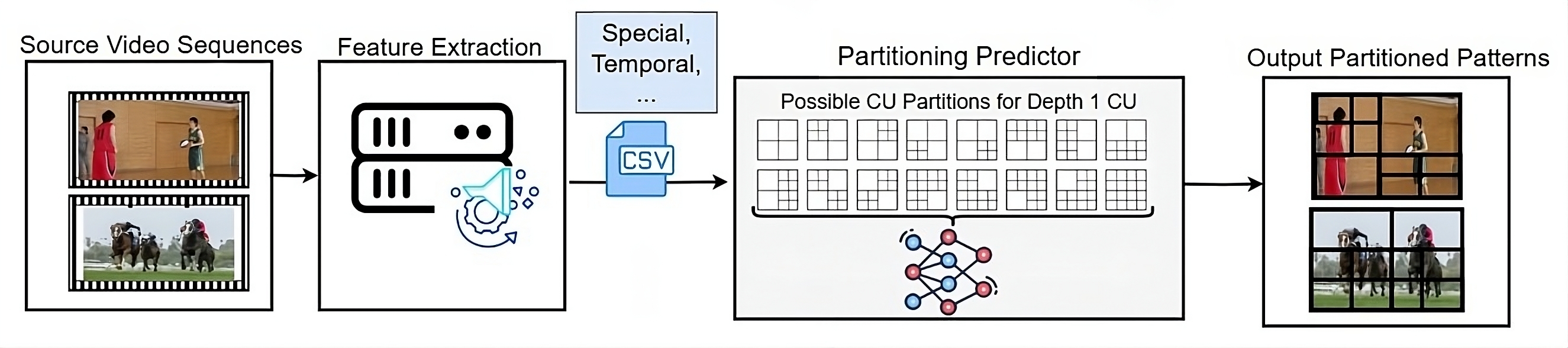}
    \caption{AI-based block partitioning example in HEVC codec.}
    \label{fig:cu}
\end{figure}

As part of the predictive coding process, it segments video frames into codec-specific block sizes that play a pivotal role in video compression. For example, in the HEVC standard, these blocks called \textit{Coding Tree Units} (CTU) have dimensions of ($16 \times 16$), ($32 \times 32$), and ($64 \times 64$), further broken down into \textit{Coding Units} (CUs) and in turn subdivided into smaller \textit{Prediction Units} (PU)~\cite{ccetinkaya2021ctu}. The intra-frame prediction uses encoded samples within the same video frame to estimate the pixels within each PU. 
Fig.~\ref{fig:cu} illustrates an example of the HEVC CU partitioning process, representing an AI-based classification task that analyzes the input video frame to extract the most relevant features, such as temporal, spatial, and motion information, fed to an AI-based classifier. The classifier determines whether to split a specific CU level based on the similarity of each region's features.
Various intra-prediction modes, \eg horizontal, vertical, and angular~\cite{nair2019machine, laude2016deep}, can identify the optimized mode with the minimum error~\cite{sullivan2012overview, zhang2017effective}. In all studies, decisions, such as whether to split a coding unit~\cite{mercat2018machine}, determine the size of the block~\cite{zhang2017effective}, and select the intra-prediction mode~\cite{jillani2020multi} {constitute integral aspects of intra-frame prediction.}

Zhang~\etal~\cite{zhang2017effective} proposed an ML model for data-driven CU size decisions to address the challenges of encoding complexity and power consumption of HEVC intra-frame prediction. Their model consists of an SVM classifier {to determine whether to split the} CU or propose an uncertain condition for situations requiring further investigation, {aiding in the prevention of false predictions}. {During instances of uncertainty}, a second stage of binary classification is performed, {leveraging insights from previously coded frames to enhance decision-making}. The results show an {average reduction in computation complexity of \qty{52.48}{\percent} compared to the original HM \num{16.7} encoder, resulting in significant energy savings}. 

The authors in~\cite{mercat2018machine} proposed an ML method to determine {the optimal partitioning of a given CU} and to predict whether a CU can be combined with the CU of the previous depth or split into more CUs. The results of their approach demonstrated that the features extracted from the depth and variance information are the most relevant parameters for predicting the CU decomposition using ML to achieve a balance between computational complexity and accuracy. 
\cite{jillani2020multi} introduced an unsupervised multi-view clustering technique to identify correlations between intra-prediction mode decisions and information extracted from the texture data of adjacent CUs in HEVC. Their approach achieved an average time reduction of around \qty{50.4}{\percent}, resulting in reduced energy consumption compared to the reference HM 16.8 encoder. 
\subsubsection{Inter-frame prediction}
\label{interframe}
Inter-frame prediction relies on previously encoded reference frames to estimate the movement of objects between frames using motion vectors and compensation. These vectors indicate the displacement of each block between consecutive frames~\cite{sullivan2012overview}. 
In~\cite{zhang2015machine}, the authors proposed a close method~\cite{zhang2017effective} using a joint SVM classifier with three outputs to determine whether to split a CU. In uncertain cases, full rate-distortion optimization determines the CU depth. This process explores all combinations and selects those that reduce the computational complexity by \qty{51.45}{\percent} compared to the original HEVC, consequently reducing the energy consumption.
\subsubsection{Motion compensation}
\label{motion_compensation}
Arnaudov~\etal~\cite{arnaudov2020artificially} proposed an ANN model for motion estimation in HD video sequences, improving video quality and balancing performance and power consumption. A dictionary of previously learned fixed search patterns and a pre-trained ANN model can select the best search pattern according to the motion dynamics within a specific region of the video frame. The method achieves the highest motion estimation quality per unit of consumed power,~\ie \qty{0.5}{\decibel} less PSNR than a complete search method while minimizing the search spacee.
\subsection{Video transcoding}
\label{transcoding}
Video transcoding involves modifying an encoded video to convert it from one coding standard to another. Another form of transcoding is adapting specific parameters like QP, bitrate, or resolution without changing the codec, often known as transrating. This process encompasses both the encoding and decoding procedures, making it a computationally complex and power-intensive operation. The state-of-the-art AI-enabled energy-aware transcoding algorithms in the literature can be categorized based on the conversion of codec, bitrate, and QP.
\subsubsection{Codec}
\label{codec}
Yang~\etal~\cite{yang2022machine} proposed an ML-based fast transcoding method from Distributed Video Coding (DVC)~\cite{lu2019dvc} to HEVC. Initial features, such as motion information extracted from the DVC decoding, capture relevant characteristics to predict the CU partition and PU modes in the HEVC encoding using an SVM binary classification task, resulting in a noteworthy \qty{57.64}{\percent} reduction in encoding complexity and energy consumption.
\subsubsection{Bitrate}
\label{bitrate}
Bubolz~\etal~\cite{bubolz2019quality}
used a Decision Tree (DT) model with \num{15} features from the decoding process, such as QP, CU position in X and Y directions, and color information to predict CU depth for the HEVC standard. This method achieved early termination in CU partitioning, leading to a \qty{49.5}{\percent} reduction in energy consumption, with only a \qty{0.66}{\percent} increase in Bjøntegaard Delta (BD)-rate utilized to compare the compression efficiency of different codecs, where positive values indicate that the new codec or configuration is less efficient in terms of bitrate or quality~\cite{barman2024bjontegaard}. 
\subsubsection{QP} 
\label{qp}
Costero~\etal~\cite{costero2019mamut} employed an RL-based approach to manage the runtime of QoS-sensitive real-time video transcoding using three agents, with the primary agent tasked with adapting the QP of the HEVC encoder. The authors strategically decomposed the design space, encompassing various encoding parameters, system settings, and resource allocation choices affecting QoS, power consumption, throughput, and video compression efficiency during transcoding,  enabling RL agents to concentrate on specific parameter subsets or configurations. Compared to equivalent approaches, the method demonstrated a notable \qty{24}{\percent} reduction in energy.
\begin{table*}
\centering
\caption{Summary of important works in video encoding (\ref{subsec:encode}) (SI: Spatial Information, TI: Temporal Information, CI: Color Information, BI: Brightness Information, MI: Motion Information, BR: Bitrate, FR: Framerate, Res: Resolution, RD: Rate-Distortion).}

\label{tab:encoding}
\resizebox{\textwidth}{!}{
\fontsize{7pt}{7pt}\selectfont
\renewcommand{\arraystretch}{1.2}
    \begin{tabular}{|c|c| c | c |c| c| c|c | c|}  
    \toprule
    
   \parbox[t]{2mm}{\multirow{3}{*}{\rotatebox[origin=c]{90}{\emph{Section}}}} &\multirow{2}{*}{\textit{Work}} & \multirow{2}{*}{\textit{Year}} &  \multirow{2}{*}{\textit{Scope}} &  \multicolumn{3}{c|}{\textit{AI Approach}} &  \multicolumn{2}{c|}{\textit{Evaluation Method}} \\ 
     \cline{5-9} 
    &&&&  \multirow{1}{*}{\textit{Input}} & \multirow{1}{*}{\textit{Model}} & \multirow{1}{*}{\textit{Output}} &   \multirow{1}{*}{\textit{Reference Software}} & \multirow{1}{*}{\textit{Dataset}}  \rule{0pt}{4.5ex}\\
    \toprule    
                          
    \parbox[t]{2mm}{\multirow{7}{*}{\rotatebox[origin=c]{90}{\ref{pertitle}}}}&\cite{katsenou2021efficient} & 2021 &  QP selection & SI, TI  & GP & QP  &  HM\cite{hm} & \cite{dataset100} \\ 
    \cline{2-9}  
    
    &\multirow{2}{*}{\cite{mueller2022context}} & \multirow{2}{*}{2022} & Bitrate ladder  & Video size,    & CNN, ANN,  & \multirow{2}{*}{Bitrate ladder} & \multirow{2}{*}{\cite{mueller2022context}}  & \multirow{2}{*}{Generated by authors} \\
    && & construction& BR, FR, Res   &  XGBoost, &  &      &   \\
   \cline{2-9}

    &\cite{amirpour2023optimizing} & 2023 & Preset selection  &  TI, BI   & RF & Video preset &  x265~\cite{x265}  & \cite{videodataset} \\
    \cline{2-9}

    &\cite{menon2024energy} & 2024 & Res selection & SI, TI, BR & RF & Encoding time, XPSNR & VVenc\cite{vvenc} & \cite{stergiou2022adapool} \\ 
   \cline{2-9} 

    & \multirow{2}{*}{\cite{qomex2024}} & \multirow{2}{*}{2024} & Res and QP  & SI, TI, FR, Res,  & \multirow{2}{*}{XGBoost} & \multirow{2}{*}{Energy and VMAF} & x264~\cite{x264}, x265~\cite{x265},  & \multirow{2}{*}{\cite{amirpour_mvcd_2024}} \\ 
    &  &  &  selection &  QP, Codec type &  &  &  VVenc\cite{vvenc}, AV1~\cite{av12021} &  \\

    \hline

    \parbox[t]{2mm}{\multirow{2}{*}{\rotatebox[origin=c]{90}{\ref{perscene}}}}&\multirow{2}{*}{\cite{seeliger2022green}} & \multirow{2}{*}{2022} & Bitrate ladder  &  \multirow{2}{*}{Video frame, Res, CI} &  CNN, ANN, &  \multirow{2}{*}{BR}  & \multirow{2}{*}{Elemental \cite{elemental}, \cite{mueller2022context}} & \multirow{2}{*}{Generated by authors} \\
    &&  & construction &  & XGBoost &    &  &  \\
    
    \hline

    \parbox[t]{2mm}{\multirow{6}{*}{\rotatebox[origin=c]{90}{\ref{redundancy}}}}&\multirow{1}{*}{\cite{zhang2017effective}} & \multirow{1}{*}{2017} & \multirow{5}{*}{CU partitioning} &  \multirow{1}{*}{CU information}  & \multirow{1}{*}{SVM} & \multirow{1}{*}{CU size} & \multirow{1}{*}{HM\cite{hm}}  & \multirow{1}{*}{\cite{bossen2013common}} \\
     
   \cline{2-3} \cline{5-9}
    
    &\multirow{1}{*}{\cite{mercat2018machine}} & \multirow{1}{*}{2018} &  & \multirow{1}{*}{CU information} & \multirow{1}{*}{ML} & \multirow{1}{*}{CU size} & \multirow{1}{*}{Kvazaar~\cite{viitanen2015kvazaar}} & \multirow{1}{*}{Generated by authors} \\
    
 \cline{2-3} \cline{5-9}
    
    &\multirow{1}{*}{\cite{jillani2020multi}} & \multirow{1}{*}{2020} &  & \multirow{1}{*}{Texture}  & \multirow{1}{*}{ML} & \multirow{1}{*}{Intra-prediction modes} &  \multirow{7}{*}{HM\cite{hm}} & \multirow{1}{*}{Generated by authors}  \\
                 
     \cline{2-3} \cline{5-7} \cline{9-9}

    &\multirow{2}{*}{\cite{zhang2015machine}} & \multirow{2}{*}{2015} &  & MI, RD and Depth    & \multirow{2}{*}{SVM} & \multirow{2}{*}{CU size} &   &  \multirow{2}{*}{\cite{bossen2013common}} \\
     &&&   & of CU, QP& &    &    &   \\
       \cline{2-7} \cline{9-9}

    &\cite{arnaudov2020artificially} & 2020 & Motion vector & MI & ANN & Searching pattern &   & \cite{xiph} \\
  \cline{1-7} \cline{9-9}
    
     \parbox[t]{2mm}{\multirow{7}{*}{\rotatebox[origin=c]{90}{\ref{transcoding}}}}&\multirow{2}{*}{\cite{yang2022machine}} & \multirow{2}{*}{2022} &  & MI, Residual  & \multirow{2}{*}{SVM} & \multirow{2}{*}{CU size} &    & \multirow{2}{*}{Generated by authors} \\
    &&  &   &   variability information &  &     & &  \\
     \cline{2-3} \cline{5-7} \cline{9-9}

   &\cite{bubolz2019quality} & 2019 & Transcoding & QP, CI, CU position  & DT & CU size &  & \cite{sharman2017common} \\
  \cline{2-3} \cline{5-9}

   & \multirow{3}{*}{\cite{costero2019mamut}} & \multirow{3}{*}{2019} & acceleration &  BR, PSNR,  & \multirow{3}{*}{RL} &  \multirow{3}{*}{QP}  &   \multirow{3}{*}{Kvazaar~\cite{viitanen2015kvazaar}}  & \multirow{3}{*}{\cite{bossen2013common}} \\    
  & &  & & Throughput, &  &      &     &  \\
   &&  & & Power consumption&  &      &     &  \\
    
    \bottomrule 
    \end{tabular}  }  
\end{table*}
\subsection{Encoding summary} 
Table~\ref{tab:encoding} summarizes the 
AI-based for encoding configuration, redundancy reduction, transcoding improvement, and considering energy consumption. XGBoost and RF are the most frequent models for regression tasks, such as bitrate or encoding time prediction, while SVM models are predominantly used for classification, particularly for CU partitioning.
Common implementations for HEVC include x265~\cite{x265}, HM~\cite{hm}, and Kvazaar~\cite{viitanen2015kvazaar}, while typical implementations for VVC are VTM~\cite{vtm} and VVenC~\cite{vvenc}.
\section{Delivery Network}
\label{subsec:delivery}
\begin{figure}[t]
    \centering
    \includegraphics[width=\linewidth]{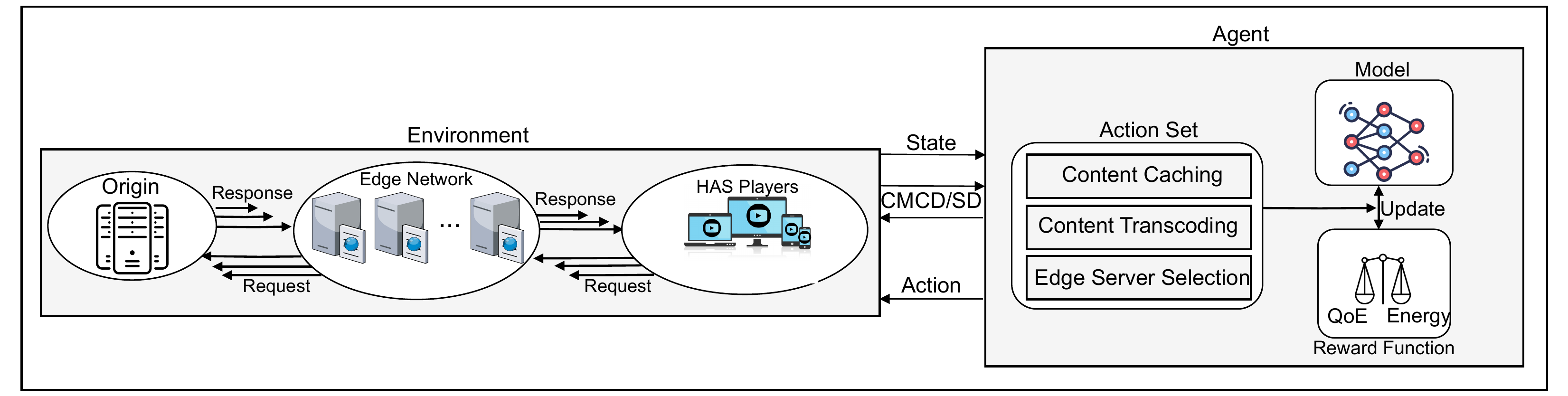}
    \caption{RL-based streaming system example.}
    \label{fig:fig7}
\end{figure}

As mentioned above, the CMCD and CMSD specifications have recently garnered attention for facilitating communication between clients and media servers. CMCD provides a standardized method for the streaming client to communicate metrics and playback data to the servers, such as buffer information, playback rate, and content identification, ensuring that the CDN can consistently interpret and analyze the data from various streaming clients. Hence, numerous AI-based network-assisted video streaming systems leverage these protocols to improve network QoS, user QoE, and energy efficiency.
One suitable AI technique for such systems is RL, which can be employed on CDN or MEC servers. 
For example, RL agents illustrated in Fig.~\ref{fig:fig7} leverage CMCD and CMSD information to incorporate network, video, user, and CDN data in their decision-making process for content cache placement~\cite{li2019deep,tang2019energy} to achieve lower latency, higher QoE, and lower delivery energy, creating a reward function communicated to the CDN or MEC servers through the CMCD system. In the following, we categorize AI-based energy-efficient network-assisted video streaming solutions according to their delivery networks. 
As several studies have introduced novel techniques for edge instances without distinguishing between MEC and CDN, we employ the \textit{edge network} term in our taxonomy.
\subsection{Edge network}
\label{edge_network}
Enhancing video streaming workflows on edge networks, whether CDN or MEC, requires pivotal choices regarding content caching, video processing task offloading and scheduling, and resource allocation methods.
\subsubsection{Content caching} \label{edge_content_caching}
Tanzil~\etal~\cite{tanzil2017adaptive} proposed a single hidden-layer feed-forward neural network, named Extreme Learning Machine (ELM), to predict the popularity of video content based on user behavior and network properties (\eg cache size, bandwidth). They used thumbnails, titles, keywords, and descriptions of YouTube video sequences as inputs to the model and proposed a Mixed Integer Linear Programming (MILP) optimization model~\cite{genova2011linear} to determine the cache with the minimum cache deployment cost, i.e., the energy consumption required to read/write files from hardware units. This method improved the cache hit ratio of \num{0.9} compared to random cache deployment, consequently improving deployment cost.

Li~\etal~\cite{li2021collaborative} presented a Multi-Layer Perceptron (MLP) model to predict user behavior based on the titles, categories, and descriptions of video sequences. They used an optimization model~\cite{quesada1992lp} to determine the optimal caching strategy for popular video content on edge servers driven by latency and energy consumption considerations, leading to an energy reduction of \qty{1800}{\joule}, a latency decrease of \qty{120}{\second}, and an \qty{80}{\percent} improvement in cache hit rates compared to baseline approaches.

Zheng~\etal~\cite{zheng2021hybrid} proposed an LSTM model Moreover, thet popular content, integrated into a Deep Deterministic Policy Gradient (DDPG)~\cite{tan2021reinforcement} algorithm to learn an optimal caching policy in energy consumption and latency in continuous VR action spaces based on user requests. The results indicated that the method exceeds baseline random selection and a traditional DDPG (without LSTM) to trade off energy and latency.

Xie~\cite{xie2024deep} optimized energy-efficient video caching and delivery in cellular networks using Deep Reinforcement Learning (DRL) agents leveraging network conditions and user demands to decide caching resources and user associations with base stations. They adjusted the model's rewards, achieving a \qty{40}{\percent} improvement in energy consumption, including the energy costs of video transmission, compared to baseline random selection.  
\subsubsection{Offloading, scheduling, and allocation} \label{edge_offloading}
Chen and Liu~\cite{chen2021energy} introduced an energy-efficient DRL-based model for video processing task offloading and resource allocation in AR applications. The state of the DRL model encompasses the data size, computing and transmission resources, and network conditions. The reward is the energy consumed when task completion and delay constraints are met. They used two methods for task offloading: binary and partial. In the binary-based method, the decision is made whether to offload a task to the edge server or execute it on the user's device. In contrast, in partial task offloading, each task is segmented into subtasks, with each subtask partially offloaded to edge servers. These subtasks include receiving the raw video frame from the camera, tracking the user's position in the environment, identifying objects, and preparing the processed frame for display. Their solution achieved an energy consumption reduction of \qty{75}{\percent} compared to a greedy approach and \qty{30}{\percent} compared to a random baseline method.

Lu~\etal~\cite{lu2020edge} introduced a QoE-aware video processing offloading method utilizing a DRL model tailored for Internet of Things (IoT) devices. The reward function of their approach incorporates service latency, energy consumption, and task success rate, while the model states encompass data size for each task, Transmission loss, and the computation resources available on edge servers. The results indicated an improvement in energy consumption and service latency by \qty{13.97}{\percent} and \qty{11.56}{\percent} compared to a DDPG approach. 
Baker~\etal in~\cite{baker2024editors} utilized an LSTM model considering energy consumption, latency, and memory usage across various tasks, particularly on video processing tasks, to identify the most reliable edge node for offloading video processing tasks such as real-time video analysis. Subsequently, they introduced a DRL model to intelligently allocate video processing tasks to the selected edge devices, aiming to expedite task execution and improve latency and energy consumption. The results demonstrated notable improvements in both energy efficiency and latency compared to baseline methods~\cite{yan2021deep, huang2019deep}.
Zhu~\etal~\cite{zhu2022energy} introduced an RL-based model to implement an energy-efficient edge resource allocation for video processing tasks. In this solution, the agent states include edge capacity and video processing task size, while the actions involve assigning edge instances to video processing tasks, specifically for handling AI computations such as training and inference. The reward function is also designed based on power consumption, providing a comprehensive energy efficiency measure within the edge computing environment. The results indicated that the energy consumption of the proposed method is \qty{30}{\percent} lower than that of the first-in, first-out baseline allocation strategy.
\subsection{Peer-to-Peer (P2P)} 
\label{p2p}
AI-based solutions are commonly applied in P2P video delivery, specifically for predicting user locations and content popularity and optimizing content caching and delivery. Li~\etal~\cite{li2019deep} employed two distinct recurrent neural network-based architectures, \ie Echo State Network (ESN)~\cite{jaeger2007echo} and LSTM, for predicting user locations and content popularity, respectively. The ESN model used historical location information, while the LSTM incorporated user-related details such as gender, age, occupation, and device type. They also introduced a DRL model to strike a balance between energy consumption and delay and to determine the optimal device for connection based on predicted data from neural networks. The proposed method achieves lower energy consumption and reduced content delivery delay compared to the baselines used in the research. 

P2P content delivery providers employ several incentives to motivate peers to share their network or computational resources to benefit the entire network. These incentives can vary according to the policies of each P2P provider. For example, the ``tit-for-tat'' principle can be offered, where active peers receive faster download speeds when they contribute their upload bandwidth to other peers. Another example is the ``reward-based'' policy, where active peers receive compensation in terms of monetary rewards or free Internet services for their resource contributions~\cite{he2022blockchain}. Therefore, to enhance video delivery and reduce server energy consumption, users can allocate a portion of their available computing resources, in addition to network bandwidth, to have distributed video processing (\eg transcoding or super-resolution) over the P2P network.

Liu~\etal~\cite{liu2018deep} introduced a DRL-based model to select optimal peers for running transcoding functions. The reward function incorporates factors such as other users' trust, historical records of successful transcoding ratios, the status of the peer battery, and its energy consumption pattern. The results validated that their approach achieved superior transcoding efficiency while meeting QoS requirements and outperformed the baseline greedy method.
Farahani~\etal~\cite{farahani2022hybrid} employed a Self Organizing Map (SOM)~\cite{kohonen-som} to propose an online learning model for handling live video requests in a hybrid P2P-CDN network. This model effectively determines the serving action, whether from the P2P side through video transcoding or caching or from the CDN side through edge caching or edge transcoding, while taking serving latency into consideration.
They extended their approach in~\cite{farahani2023alive} by incorporating a video super-resolution running at the client side called ``peer super-resolution'' and considering delivery cost in their decision-making process. The results revealed a \qty{22}{\percent} improvement in users' QoE, a \qty{34}{\percent} reduction in streaming service provider costs, and a \qty{31}{\percent} decrease in edge server energy consumption compared to alternative greedy strategies.
\begin{table*}
\captionsetup{skip=10pt}
\caption{Summary of important works in delivery network (\ref{subsec:delivery}) (RTT: Round Trip Time, Bandwidth: BW).}
\label{tab:delivery}
\resizebox{\textwidth}{!}{
\fontsize{6.9pt}{6.9pt}\selectfont
\centering    
\renewcommand{\arraystretch}{1.2}
     \begin{tabular}{|c|c| c | c |c |c| c|c | c|}  
    \toprule
    
    \parbox[t]{2mm}{\multirow{3}{*}{\rotatebox[origin=c]{90}{\emph{Section}}}}&\multirow{2}{*}{\textit{Work}} & \multirow{2}{*}{\textit{Year}} &  \multirow{2}{*}{\textit{Scope}} &  \multicolumn{3}{c|}{\textit{AI Approach}} &  \multicolumn{2}{c|}{\textit{Evaluation Method}} \\ 
    \cline{5-9} 
    
    &&&&  \multirow{1}{*}{\textit{Input}} & \multirow{1}{*}{\textit{Model}} & \multirow{1}{*}{\textit{Output}} &   \multirow{1}{*}{\textit{Real Testbed/Simulator}} & \multirow{1}{*}{\textit{Dataset}}  \rule{0pt}{4.5ex}\\
    \toprule    
  
    \parbox[t]{2mm}{\multirow{9}{*}{\rotatebox[origin=c]{90}{\ref{edge_content_caching}}}}&\multirow{2}{*}{\cite{tanzil2017adaptive}} & \multirow{2}{*}{2017} & \multirow{9}{*}{Cache placement} &  Thumbnails, Title,   & \multirow{2}{*}{ELM} & \multirow{5}{*}{Popular video} &  \multirow{2}{*}{NS-3\cite{mastorakis2016ndnsim}} & \multirow{2}{*}{Generated by authors}  \\
     &&&  &   Keywords, Description &  & &      & \\
     \cline{2-3} \cline{5-6} \cline{8-9} 

    &\multirow{3}{*}{\cite{li2021collaborative}} & \multirow{3}{*}{2021} &  &  Video title, Category,  & \multirow{3}{*}{MLP} &  &  \multirow{3}{*}{ \num{10} servers, \num{25} clients } & \multirow{3}{*}{\cite{wolf2011face, youtubeface}} \\
    & &  &  & Streaming duration,  & &    &      &  \\
    & &  &  & Description, No. requests & &  &  &  \\      
     \cline{2-3} \cline{5-9} 

    &\multirow{2}{*}{\cite{zheng2021hybrid}} & \multirow{2}{*}{2021} &  &  Cache capacity,    & \multirow{4}{*}{DRL} & \multirow{4}{*}{Cache location} &  \multirow{2}{*}{Numerical simulation} & \multirow{2}{*}{Generated by authors} \\
    & & &  &  Video popularity   &  &  &   &  \\
     
   \cline{2-3} \cline{5-5} \cline{8-9} 

    &\multirow{2}{*}{\cite{xie2024deep}} & \multirow{2}{*}{2024} &  & Cache capacity, Network   & &  & Numerical simulation & \multirow{2}{*}{Generated by authors} \\
    && & & BW and congestion  & & & (\num{10} clients) & \\
    
    \hline 
    
    
     \parbox[t]{2mm}{\multirow{9}{*}{\rotatebox[origin=c]{90}{\ref{edge_offloading}}}}&\multirow{3}{*}{\cite{chen2021energy}} &  \multirow{3}{*}{2021} & Video processing  &  Task size, &  \multirow{8}{*}{DRL} &  \multirow{8}{*}{Offloading location}   &  Numerical simulation   &  \multirow{3}{*}{Generated by authors}  \\ 
    &&  & task offloading and &  Network BW,  &  &   & (\num{3} servers, \num{15} clients,    &   \\
     &&  & resource allocation & Computing capacity  &  &   & \num{3} base stations)    &   \\
   \cline{2-5} \cline{8-9} 

    &\multirow{2}{*}{\cite{lu2020edge}} & \multirow{2}{*}{2020} & Video processing & Task size, Transmission  & &  &   \multirow{2}{*}{Numerical simulation}   & \multirow{2}{*}{Generated by authors} \\
         &&  & task offloading&    loss, Computational capacity &  & &     & \\ 
     \cline{2-5} \cline{8-9}

    &\multirow{3}{*}{\cite{baker2024editors}} & \multirow{3}{*}{2024} & Video processing & Energy consumption, Execution &  &  & Python simulation & \multirow{3}{*}{\cite{dos2021algorithm, deng2014pedestrian}} \\
     &&  & task & and communication time,   &  &  & (\num{5} edge devices, \num{10} tasks) &  \\
      &&  & offloading & BW availability, Trust scores &  &  &  &  \\    
   \cline{2-9} 

    &\multirow{2}{*}{\cite{zhu2022energy}} & \multirow{2}{*}{2022} & Video processing & Computational capacity,  & \multirow{2}{*}{RL} & \multirow{2}{*}{Resource scheduling} &   \num{200} NVIDIA & \multirow{2}{*}{Generated by authors} \\ 
     &&& resource allocation & Task size  &  &  &   Jeston devices \cite{jetson}&  \\ 
       
    \hline
    \parbox[t]{2mm}{\multirow{6}{*}{\rotatebox[origin=c]{90}{\ref{p2p}}}}&\multirow{3}{*}{\cite{li2019deep}} & \multirow{3}{*}{2019} & \multirow{3}{*}{Cache placement} &  User's information,  & ESN,  & Popular video,  &  Numerical simulation & \multirow{3}{*}{Generated by authors} \\ 
     &&  &  &   Age, Location & LSTM &  Device location &  (\num{200} peers)    &  \\ 
   \cline{5-7} 
    
    & &  &  &  \multirow{1}{*}{Device location, Popularity}& \multirow{1}{*}{DRL} & \multirow{1}{*}{Cache location} &   & \\
    
   \cline{2-9} 
    &\multirow{2}{*}{\cite{liu2018deep}} & \multirow{2}{*}{2018}  & Peer selection  &  Workload, Energy consumption,   & \multirow{2}{*}{DRL} & \multirow{2}{*}{Transcoder selection} &  Numerical simulation   & \multirow{2}{*}{Generated by authors} \\
     &&  & for transcoding & QoS, peer's trust score &  &     & (\num{10} peers) & \\
   \cline{2-9}

      &\multirow{3}{*}{\cite{farahani2022hybrid, farahani2023alive}} & \multirow{3}{*}{2022} & Request serving & Buffer status, BW,   & \multirow{3}{*}{SOM} & \multirow{3}{*}{Serving action} & \multirow{3}{*}{\num{5} servers, \num{350} peers} & \multirow{3}{*}{\cite{lederer2012dynamic,zoo}} \\
    &&  & by super-resolution  &  Bitrate,  &  &  & &  \\ 
     &&  &  and transcoding & Computation power  &  &  & &  \\ 
         \hline

    \parbox[t]{2mm}{\multirow{4}{*}{\rotatebox[origin=c]{90}{\ref{emerging}}}}&\multirow{2}{*}{\cite{zhong2022q}} & \multirow{2}{*}{2022} & MPTCP delivery &  Packet loss, RTT, & \multirow{2}{*}{RL} & \multirow{2}{*}{Transmission path}& \multirow{2}{*}{NS-3\cite{ns3}}  & \multirow{2}{*}{Generated by authors} \\
     &&  &path selection &  Energy consumption,  &  &&    & \\
   \cline{2-9}   

    &\multirow{2}{*}{\cite{shams2023managing}} & \multirow{2}{*}{2023} & Network selection & Network throughput,   & \multirow{2}{*}{MLP} & \multirow{2}{*}{Connection device} &  \multirow{2}{*}{\num{10} Raspberry Pis} & \multirow{2}{*}{Generated by iperf \cite{iperf}} \\
    && & for device connection &  Energy consumption  & & &   &  \\
    \bottomrule
    
    \end{tabular}  }
\end{table*}
\subsection{Emerging delivery protocols}
\label{emerging}
In recent years, various networking protocols, such as QUIC~\cite{QUIC}, Web Real-Time Communication (WebRTC)~\cite{WebRTC}, and Multipath Transmission Control Protocol (MPTCP)~\cite{ford2013tcp}, have been utilized in the streaming domain~\cite{mohamed2023survey, zverev2021robust, ran2023isaw, han2024multi}. These protocols contribute to reducing latency and QoE. 
Our search methodology revealed a limited emphasis on leveraging AI models to enhance energy efficiency and QoE within such protocols. For instance, the authors in \cite{zhong2022q} introduced an RL-based model for MPTCP video delivery. The RL agent determines when and over which path the video packets should be transmitted, optimizing the delivery of high-quality 5G media services. The state of the proposed RL model encompasses the packet loss rate in each path, the energy consumption of data transferring in each path, and the Round-Trip Time (RTT), while the reward function is composed of throughput, packet loss, and energy consumption. This method outperformed the lowest-RTT-first~\cite{paasch2014experimental} and earliest completion first~\cite{lim2017ecf} baseline methods in terms of completion time, transmission rate, and energy consumption. 
Shams~\etal~\cite{shams2023managing} proposed a Deep Learning (DL) model to improve the energy consumption of MPTCP video delivery using two algorithms: a Software-Defined Network (SDN)-based method determining the network for device connection; an algorithm adjusting the decision of the first algorithm based on energy consumption. They also utilized an MLP model pre-trained on different network sizes and MPTCP congestion control algorithms. This method reduced energy consumption by \qty{19}{\percent} and improved network performance by \qty{8}{\percent} compared to other baselines, such as MPTCP Cubic~\cite{rhee2018cubic}.
\subsection{Delivery summary} 
Table~\ref{tab:delivery} summarizes works that utilized AI techniques to present energy-aware delivery services provided by single or hybrid edge-P2P  networks as well as the features of emerging delivery protocols. Our analysis shows that RL-based methods, particularly DRL, are the most popular models for offloading, scheduling, allocation, and caching problems. Moreover, the most common evaluation approaches are simulating the end-to-end aspects of video streaming by numerical simulations, Python-based frameworks, or NS-3.

\section{Video Playback}
\label{subsec:player}
In recent years, AI techniques have enhanced ABR algorithms, incorporating energy efficiency alongside traditional factors like QoE and network throughput.
In addition, client-side AI-based video quality enhancement techniques, such as video super-resolution and contrast adjustments, are typically used to optimize QoE during video playback.
\subsection{ABR algorithms}
\label{abr}
Turkkan~\etal~\cite{turkkan2022greenabr} developed an RL-based ABR algorithm called GreenABR, incorporating states of energy consumption, network throughput, video player current buffer size, the bitrate of the most recently downloaded segment, and the VMAF score. They used an ANN model to predict the playback energy consumption with only \qty{7}{\percent} error based on input variables such as bitrate, resolution, framerate, file size, and VMAF. The ANN-based energy model, leveraging network throughput and video player measurements, optimizes the selection of video representations, ensuring energy-efficient choices. They designed the RL reward function based on quality, buffering, and energy consumption factors. The results demonstrated a \SI{57}{\percent} reduction in power consumption compared to the buffer-based BOLA ABR algorithm~\cite{spiteri2020bola}.    
The authors extended their research to GreenABR+~\cite{turkkan2024greenabr+}, employing the same energy model and using VMAF as the perceived quality metric. GreenABR+ tackles the challenge of retraining models for diverse video representation sets by employing a DRL model encompassing a broader bitrate range. GreenABR+ outperforms both GreenABR~\cite{turkkan2022greenabr} and Pensieve~\cite{mao2017neural}, demonstrating a \qty{57}{\percent} reduction in energy consumption and an \qty{87}{\percent} improvement in rebuffering time.

Mondal~\etal~\cite{mondal2020endash} introduced an energy-efficient DASH video player (EnDASH) designed to balance user QoE and energy consumption for devices in mobility scenarios. The player employed three consecutive prediction models, resulting in a nearly \SI{30}{\percent} reduction in energy consumption compared to the state-of-the-art Pensive~\cite{mao2017neural}:
\begin{enumerate*}
    \item an \textit{RF model} to predict the average throughput of the cellular network using historical information and characteristics such as associated cellular technology (\ie 2G-4G) and the number of neighboring base stations;
    \item an \textit{RL model} to predict and determine whether to increase or decrease the buffer length based on a reward function considering energy savings and QoE score. The RL states include the average cellular network throughputs predicted in the previous step, the current playback buffer capacity, and a set of potential changes in the buffer length;
    \item a \textit{DRL model} utilizes states, such as the predicted buffer length, throughput, and the previous segment's bitrate, to select the next segment based on the QoE score.
\end{enumerate*} 

Given the belief in the necessity for an ABR algorithm to adapt more effectively to specific network conditions rather than performing well on average across all conditions, Huang~\cite{huang2022learning} proposed an ABR algorithm, $A^{\scriptsize 2}BR$, consisting of online and offline stages. In the offline stage, they trained an RL model with diverse scenarios of network conditions. This trained model is then utilized by the video player in the online stage to select the optimal bitrate based on the current network conditions. The results confirmed a \SI{7}{\percent} improvement in VMAF while consuming the same amount of energy compared to the baseline approach~\cite{yin2015control}.
Raman~\etal~\cite{raman2024ll} introduced a DRL method for energy-efficient live video streaming on mobile devices called LL-GABR. LL-GABR states encompassed player metrics such as current buffer size, latency, rebuffering time, playback rate, bitrate, VMAF, network predicted bandwidth and player energy consumption. They designed a reward function based on high video quality, minimized rebuffering, reduced bitrate switches, and enhanced energy efficiency. LL-GABR showed performance improvements compared to other adaptive bitrate algorithms, such as \cite{gutterman2020stallion, sun2021tightrope}, achieving a \qty{44}{\percent} improvement in QoE and a \qty{73}{\percent} enhancement in energy efficiency.
\subsection{Video super-resolution}
\label{sr}
Super-resolution refers to a computational task to upscale and restore high-frequency signals that may be lost in a low-resolution image or video, ultimately generating a high-resolution counterpart. This process aims to minimize visual degradation and enhance image quality as much as possible, resulting in a high-resolution image or video~\cite{wang2020deep}. 
Many DL-based super-resolution models such as~\cite{dong2016accelerating,ledig2017photo,ahn2018fast,shi2016real,liu2021evsrnet} have emerged in recent years that stand out for their innovative architectures and impressive performance. 

Given that video super-resolution often runs on end-user devices with limited resources and batteries, significant research efforts have been directed towards leveraging AI to design resource and energy-aware super-resolution techniques.
Menon~\etal~\cite{menon2024video} introduced a method to determine the encoding resolution by maximizing the achievable VMAF using FSRCNN~\cite{dong2016accelerating}. They employed an RF model to predict the VMAF scores of the video after applying a super-resolution model and the encoding time for each determined resolution. Their results showed a \qty{32.70}{\percent} reduction in bitrate while maintaining the same VMAF compared to the HLS bitrate ladder, leading to a \qty{68.21}{\percent} reduction in encoding energy consumption.
The authors in~\cite{yue2022eesrnet} proposed a lightweight CNN model by enhancing the efficiency of a well-performing model, ABPN~\cite{du2021anchor}, by pruning residual connections. Residual connections, commonly used in neural networks such as ABPN, facilitate the flow of information by providing shortcut paths for gradients during training. The pruning process indeed reduces computational overhead by selectively removing redundant connections within residual blocks without reducing the model's performance. The optimized model achieved a power consumption of \SI{0.7}{\watt}, an inference time of \SI{2.03}{\milli\second}, and a PSNR of \SI{27.46}{\decibel}. The results showcased lower power consumption and \qty{43}{\percent} improvement in inference time compared to ABPN while preserving nearly identical PSNR levels.

Gao~\etal~\cite{gao2022rcbsr} introduced a re-parameterization technique to enrich the network's feature extraction capabilities with multiple paths during training. Subsequently, they expedited network inference by consolidating multiple paths into a single convolution block. They achieved a PSNR of \num{27.52} with a power consumption of \SI{0.1}{\watt} on the identical dataset used in \cite{yue2022eesrnet}. Zhang~\etal~\cite{zhang2023efficient} proposed an efficient method for running Deformable Convolutional Networks (DCNs) on Field Programmable Gate Arrays (FPGAs) for video super-resolution. In their approach, the authors introduced optimizations at two levels: algorithm and hardware. In algorithm-level optimization, they used a lightweight network to reduce complexity and memory usage, while at the hardware level, they designed a dedicated processing core and storage to handle irregular memory access caused by deformable convolutions. The experiments demonstrated a \num{1.63} times enhancement in energy efficiency compared to other FPGA-based super-resolution baselines. 

The authors in~\cite{farahani2023alive} proposed a framework for live video streaming (ALIVE), considering all available computational resources from peers, MEC servers, and CDNs. They leveraged video super-resolution for DASH requests in the P2P layer of their architecture, upscaling low-resolution frames received from adjacent peers. Experimenting with various CNN-based super-resolution techniques, they selected CARN~\cite{ahn2018fast} for PC and LiDeR~\cite{ccetinkaya2022lider} for mobile peers. The framework achieved a \SI{22}{\percent} improvement in QoE and a \SI{31}{\percent} reduction in edge energy consumption. Inspired by the concept of super-resolution to construct high-resolution videos from low-resolution counterparts, the authors in~\cite{moghaddam2024nu} employed a CNN-based model to mitigate the visual distortions inherent in video encoding, adeptly and effectively reconstructing the video sequence from low-bitrate video sequences, yielding a \qty{40}{\percent} improvement in perceived quality, as measured by both PSNR and SSIM metrics Compared to baseline methods.
\subsection{Video contrast adjustment}
\label{contrast}
The user device's display, whether a smartphone, laptop, TV, or PC, constitutes another component contributing to power consumption in the video streaming lifecycle. The display's power consumption typically depends on its technology, such as LCD, OLED, or AMOLED~\cite{kim2013runtime}. Video contrast on displays involves the differentiation in brightness or color between each frame's brightest and darkest element, impacting visibility and detail sharpness. Proper contrast alignment optimizes display technology utilization and minimizes unnecessary power consumption while improving visibility without compromising energy efficiency in the display process.

Shin~\etal~\cite{shin2019unsupervised} presented a CNN model tailored for OLED displays, aiming to boost video frame contrast and decrease power consumption. Their methodology selectively reduced brightness by a specified ratio while preserving perceived quality through contrast enhancement using a conditional generative adversarial network. This network generated output frames under the constraint of power consumption, resulting in a significant power saving rate of up to \SI{42.1}{\percent}.
Meur~\etal~\cite{le2023energy} focused on reducing display energy consumption using a pixel-wise content-adaptive method to generate an attenuation map on the encoder side. They used a CNN model, providing information on adjusting pixel values for energy reduction. The results indicated that at an energy reduction rate of \qty{10}{\percent}, the average VMAF score for different video sequences demonstrated a \qty{6}{\percent} improvement compared to a linear scaling method. 
\begin{table*}
\captionsetup{skip=10pt}
\caption{Summary of important works in video playback (\ref{subsec:player}) (BR: Bitrate, BS: Base Station, RTT: Round Trip Time, SR: Super-resolution, TR: Transcoding CA: Contrast Adjustment). }
\label{tab:player}
\resizebox{\textwidth}{!}{
\fontsize{7pt}{7pt}\selectfont
\centering
\renewcommand{\arraystretch}{1.2}
    \begin{tabular}{|c|c| c | c |c| c| c|c | c|}  
    \toprule
    \parbox[t]{2mm}{\multirow{3}{*}{\rotatebox[origin=c]{90}{\emph{Section}}}}&\multirow{2}{*}{\textit{Work}} & \multirow{2}{*}{\textit{Year}} &  \multirow{2}{*}{\textit{Scope}} &  \multicolumn{3}{c|}{\textit{AI Approach}} &  \multicolumn{2}{c|}{\textit{Evaluation Method}} \\ 
    \cline{5-9} 
    &&&&  \multirow{1}{*}{\textit{Input}} & \multirow{1}{*}{\textit{Model}} & \multirow{1}{*}{\textit{Output}} &   \multirow{1}{*}{\textit{Real Device/Simulator}} & \multirow{1}{*}{\textit{Dataset}}  \rule{0pt}{4.5ex}\\
    \toprule

 \parbox[t]{2mm}{\multirow{13}{*}{\rotatebox[origin=c]{90}{\ref{abr}}}} &\multirow{4}{*}{\cite{mondal2020endash}} & \multirow{4}{*}{2020} &   &  Cellular technology (\num{2}G-\num{4}G), & \multirow{2}{*}{RF} & \multirow{2}{*}{Throughput} & Moto G5, & Generated by  \\
     &&  & Developing DASH  &  No. BSs, Download rate &  &  &  Micromax Canvas    &  authors \\
    
       \cline{5-7} 
    & & &video players &  Network throughput, Buffer   & \multirow{2}{*}{RL} & \multirow{2}{*}{Bitrate} &   Infinity   &  \\
   & &  &  &  size, Buffer length &  &  &     &  \\     
        \cline{2-9}

   &\multirow{2}{*}{\cite{turkkan2022greenabr}, \cite{turkkan2024greenabr+} } & \multirow{2}{*}{2022} &  &  Energy consumption, BR,  VMAF, & \multirow{2}{*}{RL - DRL} & \multirow{2}{*}{Bitrate} &  \multirow{2}{*}{Samsung S4 } & \multirow{2}{*}{\cite{duanmu2018quality}} \\
    &&  &  &   Network throughput, Buffer size  &  &   &  &  \\     
   \cline{2-3}  \cline{5-9}

    &\multirow{2}{*}{\cite{huang2022learning}} & \multirow{2}{*}{2022} &  &  VMAF, Network throughput,  & \multirow{2}{*}{RL} & \multirow{2}{*}{Bitrate} &   \multirow{2}{*}{ MacBook Pro}  & \multirow{2}{*}{\cite{yan2020learning},\cite{narayanan2020lumos5g}, \cite{pendata}} \\
    &&  & Developing &  Download time, Buffer occupancy&   &  &  &    \\   
      \cline{2-3}  \cline{5-9}

    &\multirow{3}{*}{\cite{bentaleb2023meta}} & \multirow{3}{*}{2023} & ABR &   Network throughput, RTT,     & \multirow{3}{*}{RL} & \multirow{3}{*}{Bitrate} &  \multirow{3}{*}{\cite{mao2019park}}    & \multirow{3}{*}{\cite{van2016http, riiser2013commute, narayanan2021variegated, duanmu2020assessing}} \\
     && & algorithms &  Playback rate, Packet loss,   & &  &    & \\ 
     && &  &   Buffer occupancy, Resolution  & &  &    & \\ 
        \cline{2-3}  \cline{5-9}

    &\multirow{2}{*}{\cite{raman2024ll}} & \multirow{2}{*}{2024} & & Buffer size, VMAF, Bitrate, Energy  & \multirow{2}{*}{DRL} & \multirow{2}{*}{Bitrate, Playback rate} & \multirow{2}{*}{Python simulation} & \multirow{2}{*}{\cite{xiph}} \\ 
    &&  & &  consumption, Rebuffering time &  &  & &  \\
    \hline 

    \parbox[t]{2mm}{\multirow{8}{*}{\rotatebox[origin=c]{90}{\ref{sr}}}}&\cite{gao2022rcbsr} & 2022 &  &  \multirow{5}{*}{Video frame} & CNN & \multirow{3}{*}{Upscaled frame} & Python simulation & \cite{reds} \\
     \cline{2-3}  \cline{6-6} \cline{8-9}

    &\multirow{1}{*}{\cite{yue2022eesrnet}}& \multirow{1}{*}{2022} &  &  & \multirow{1}{*}{CNN} &  & Redmi K50 Pro  & \multirow{1}{*}{\cite{reds}} \\
   
     \cline{2-3}  \cline{6-6} \cline{8-9}

    &\cite{zhang2023efficient} & 2023 & Developing &  & DCN &  & Python simulation & ~\cite{xue2019video} \\
    \cline{2-3}  \cline{6-9}

    &\multirow{2}{*}{\cite{moghaddam2024nu}} & \multirow{2}{*}{2024} & SR methods  &  & \multirow{2}{*}{CNN} & Higher quality  & \multirow{2}{*}{Python, NVIDIA GPUs} & \multirow{2}{*}{Generated by authors} \\ 
    && &   &  &  & (PSNR, SSIM) &  &  \\  
     \cline{2-3}  \cline{5-9}
     
    &\cite{menon2024video} & 2024 & & Spatial, temporal information & RF & Resolution & Python simulation & \cite{videodataset} \\ 
    
     \cline{2-9}    

    &\multirow{2}{*}{\cite{farahani2023alive}} & \multirow{2}{*}{2023} &SR and TR on battery- & \multirow{5}{*}{Video frame} & \multirow{5}{*}{CNN} & \multirow{2}{*}{Upscaled frame} & iPhone 11, Xiaomi   & \multirow{2}{*}{\cite{lederer2012dynamic,zoo}} \\
    & &  &constrained end devices &  &  &  &  Mi11,  virtual machines\cite{ricci2014introducing} &  \\
     \cline{1-4} \cline{7-9} 

   \parbox[t]{2mm}{\multirow{3}{*}{\rotatebox[origin=c]{90}{\ref{contrast}}}}&\cite{shin2019unsupervised} & 2019 & \multirow{3}{*}{ CA methods} & & & Higher contrast & Python simulation &  \cite{arbelaez2010contour}\\
      \cline{2-3} \cline{7-9}

    &\multirow{2}{*}{\cite{le2023energy}} & \multirow{2}{*}{2023} &  & &  &  Pixel-value  &   \multirow{2}{*}{OLED display} & \multirow{2}{*}{\cite{bossen2019jvet,arbelaez2010contour}} \\
    && &  &  &  &  modification map &   &   \\
   \bottomrule

    \end{tabular}}
    
\end{table*}
\subsection{Playback Summary}
Table~\ref{subsec:player} summarizes AI-based ABR algorithms that incorporate content-aware information to enhance bitrate selection and energy efficiency, primarily utilizing (D)RL models and evaluated on smartphones. Our investigation also highlights client-based AI technologies, i.e., super-resolution and video contrast adjustment, which aim to balance quality enhancement with energy efficiency. These works mainly used CNN-based models and employed a combination of real devices and simulations for evaluation.
\section{Video Quality  Assessment (VQA)}
\label{subsec:vqa}
Expanding on the taxonomy of VQA metrics introduced in Section~\ref{sec:background}, these metrics can be further categorized as Full Reference (FR), Reduced Reference (RR), and No Reference (NR) methods, depending on the degree of information available for the reference video~\cite{VQ_survey}. 
VMAF, PSNR, and SSIM are in the FR-VQA category since they evaluate the quality of compressed video by directly comparing it to the original version. In contrast, NR-VQA methods do not utilize the original video content for quality assessment. 
Therefore, AI techniques are frequently employed to analyze distorted videos independently without relying on a reference video. Moreover, RR-VQA methods offer the advantage of evaluating video quality using limited information, resulting in lower power consumption compared to FR-VQA methods, making them suitable for energy-efficient video streaming applications~\cite{rr_vqa_survey}.
\subsection{Reduced Reference (RR) and No Reference (NR) VQA prediction}
\label{vmaf}
VMAF is a popular choice for quality evaluation due to its stronger correlation with perceptual quality compared to SSIM and PSNR. However, its computational intensity and high time complexity pose challenges~\cite{vmaf_ref_madhu}. Menon~\etal~\cite{menon2024VQAA} introduced an LSTM-based RR-VQA approach to predict VMAF scores for reconstructed videos. Their proposed method incorporated features such as Discrete Cosine Transform (DCT)-based texture information and a fusion of SSIM between the original and reconstructed videos. The results demonstrated a \num{9.14} times acceleration compared to traditional VMAF estimation, resulting in an \SI{89.44}{\percent} reduction in energy consumption. Barman~\etal~in~\cite{barman2019no} used four learning models, GP, MLP, Support Vector Regression (SVR), and RF to propose an NR-VQA method to predict VMAF scores in video gaming applications. The models incorporated input such as encoding parameters, content, and compression artifacts besides Blind Image Quality Index (BIQI)~\cite{moorthy2010two} and Natural Image Quality Evaluator (NIQE)~\cite{mittal2012making} as two well-known NR image quality methods adapted for video sequences. The results revealed that the SVR model exhibited the best performance compared to ground-truth VMAF.
\subsection{QoE prediction}\label{qoe}
Many works consider bitrate a primary factor in QoE assessment. However, the use of HTTPS by OTT providers to secure data transmission complicates the packet inspection and information extraction (e.g., bitrate) from encrypted network traffic~\cite{duanmu2020assessing,li2023real}.
Pan and Cheng~\cite{pan2018qoe} employed three ML models: RF for handling diverse data patterns, Bayesian Networks for probabilistic reasoning, and AdaBoost for enhancing weak classifiers, to evaluate QoE and energy efficiency in encrypted YouTube streaming. They specifically trained these models to estimate video bitrates from HTTPS traffic, resulting in energy savings of up to \qty{20}{\percent}.

Ickin~\etal~\cite{ickin2021qoe} introduced a Federated Learning (FL) technique for QoE estimation, specifically focusing on energy efficiency and lightweight training. Their approach involves partitioning the dataset across various local devices, each holding a subset of features and a central model trained by communicating with these distributed datasets. Comparative analysis against a fully centralized model showcased the efficacy of their proposed method, demonstrating comparable performance. The authors delved into hyperparameter tuning, focusing on energy efficiency and light training, resulting in a noteworthy \qty{28}{\percent} reduction in training time and a subsequent decrease in energy consumption associated with model training. Kougioumtzidis~\etal~\cite{kougioumtzidis2024qoe} addressed the influence of the wireless transmission channel on user QoE and network QoS parameters, such as bandwidth, jitter, latency, and packet loss. They used a CNN-based model to predict the QoE of video games in wireless networks, promoting efficient utilization of computational resources and reducing network energy consumption. This method achieved an MAE value of \num{0.09}, outperforming baselines with MAE metrics ranging from \num{0.1} to \num{0.7}.
\begin{table*}
\captionsetup{skip=10pt}
\caption{Summary of important works in video quality assessment (\ref{subsec:vqa}) (BR: Bitrate, SI: Spatial Information, TI: Temporal Information, BI: Brightness Information, FR: Framerate).}
\label{tab:vqa}
\resizebox{\textwidth}{!}{
\renewcommand{\arraystretch}{1.2}
\fontsize{7pt}{7pt}\selectfont
\centering    
\begin{tabular}{|c|c| c | c |c| c| c|c | c|}  
\toprule
    \parbox[t]{2mm}{\multirow{3}{*}{\rotatebox[origin=c]{90}{\emph{Section}}}}&\multirow{2}{*}{\textit{Work}} & \multirow{2}{*}{\textit{Year}} &  \multirow{2}{*}{\textit{Scope}} &  \multicolumn{3}{c|}{\textit{AI Approach}} &  \multicolumn{2}{c|}{\textit{Evaluation Method}} \\ 
    \cline{5-9} 
    
    &&&&  \multirow{1}{*}{\textit{Input}} & \multirow{1}{*}{\textit{Model}} & \multirow{1}{*}{\textit{Output}} &   \multirow{1}{*}{\textit{Device/Reference Software/Simulator}} & \multirow{1}{*}{\textit{Dataset}}   \rule{0pt}{4.5ex}\\    
    \toprule

     \parbox[t]{2mm}{\multirow{4}{*}{\rotatebox[origin=c]{90}{\ref{vmaf}}}}&\cite{menon2024VQAA} & 2024 &  \multirow{3}{*}{VMAF prediction} &  SI, TI, BI & LSTM & VMAF &  x264\cite{x264} & \cite{videodataset} \\     
    \cline{2-3}   \cline{5-9}

    &\multirow{2}{*}{\cite{barman2019no}} & \multirow{2}{*}{2019} &  &  BR, Res, SI, TI, Blur, & \multirow{2}{*}{SVR} & \multirow{2}{*}{VMAF} &   \multirow{2}{*}{MATLAB simulation} & \multirow{2}{*}{\cite{barman2019no,barman2018gamingvideoset}} \\
     &&  &  &     Blockiness, BIQI,  NIQE   &    &  &  \\
        
    \hline
    
    \parbox[t]{2mm}{\multirow{5}{*}{\rotatebox[origin=c]{90}{\ref{qoe}}}}&\multirow{2}{*}{\cite{pan2018qoe}} & \multirow{2}{*}{2018} & \multirow{6}{*}{MOS prediction} &  BR, Segment size,  & RF, AdaBoost,  & \multirow{2}{*}{Bitrate} & iPhone 5, 6, 6s, HTC M7, & \multirow{2}{*}{Generated by authors} \\
        & &  &  & Res, Segment duration  & Bayes Network &  &  Samsung S4, Xiaomi MI2   &  \\
       \cline{2-3}   \cline{5-9}

    &\multirow{2}{*}{\cite{ickin2021qoe}} & \multirow{2}{*}{2021} &  &  SI, TI, FR, BR, No. stalls,  & \multirow{2}{*}{Vertical FL} & \multirow{2}{*}{MOS} &  \multirow{2}{*}{Python simulation} & \multirow{2}{*}{\cite{duanmu2018quality}} \\
    &&  &  &   Stall duration, MOS  & &    &     &  \\   \cline{2-3}   \cline{5-9}

    &\multirow{2}{*}{\cite{kougioumtzidis2024qoe}} & \multirow{2}{*}{2024} &  & Bandwidth, Jitter, & \multirow{2}{*}{CNN} & \multirow{2}{*}{MOS} &  \multirow{2}{*}{OpenAirInterface~\cite{kaltenberger2020openairinterface}} & \multirow{2}{*}{Generated by authors} \\
     & &  &  &  Network latency, Packet loss &  &  &   &  \\

   \bottomrule
    \end{tabular}}
        
\end{table*}
\subsection{VQA summary}
Table~\ref{tab:vqa} summarizes the work that utilized AI techniques to reduce the dependency on reference frames by the NR and RR-VQA methods. Moreover, it highlights work that predicts QoE information, such as bitrate, within HTTPS traffic, aiming to balance QoE assessment and energy efficiency. The investigation reveals no single dominant ML model or evaluation method for the discussed contributions. The choice of model and evaluation approach varies based on specific sub-problems.

\section{Future Research Directions}
\label{sec:challenge}
This section outlines potential future research directions for AI-driven, energy-efficient video streaming systems, focusing on the segments of the video streaming lifecycle and the role of generative AI in enhancing the performance of these segments.
\subsection{\textit{Video encoding}} 
Beyond integrating AI into conventional codecs like HEVC and VVC, there is growing interest in end-to-end learning-based coding schemes~\cite{zhang2017effective, jillani2020multi, mercat2018machine}. Unlike traditional methods that optimize encoding and decoding separately, end-to-end approaches use integrated DL models to optimize the entire video compression process as a unified learning task. In this context, extensive datasets, including pairs of original video frames and their compressed representations, are typically used to train networks, enabling them to proficiently map input video frames to their compressed counterparts~\cite{jia2023mpai}. In particular, the MPAI-EEV standard, developed by the Moving Picture, Audio, and Data Coding by Artificial Intelligence (MPAI) group, demonstrates superior perceptual evaluation metrics compared to the VVC codec~\cite{jia2023mpai}. For example, in~\cite{ma2020end}, a DL-based end-to-end compression scheme achieved a \qty{23}{\percent} improvement in BD-rate on average compared to HEVC. However, further research is needed to explore the impact of this approach on energy consumption and carbon emissions similar to the studies like~\cite{katsenou2022energy,kranzler2020comparative,herglotz2017decoding} conducted on existing codecs.
\subsection{\textit{Delivery network}} 
The increasing prevalence of edge computing, softwarized and virtualized networks, and technologies like \emph{EdgeAI} is driving the integration of AI into critical network and resource management tasks. These tasks include optimizing resource allocation, scheduling video processing, and enhancing user request servicing through intelligent server and segment selection strategies. Given the resource-intensive nature of AI models, it is imperative to optimize both accuracy and energy consumption~\cite{xu2024unleashing,ji2023e2e}. A promising strategy involves decentralizing computations, encompassing both training and inference, by leveraging edge devices. This approach offers significant advantages, particularly in reducing computation time and energy consumption~\cite{zhang2023towards}.
Leveraging emerging computing technologies, such as serverless computing~\cite{denninnart2024smse,patros2021toward}, 
with AI to enhance the sustainability, efficiency, and monetary cost of video streaming systems presents new opportunities. In addition, using intelligent in-network components to assist video encoders in reducing bitrates and energy consumption through constructing dynamic bitrate ladders while maintaining high quality and user experience, or enhancing network resource utilization in multi-CDN environments~\cite{viola2020predictive}, e.g., bandwidth and server capacity, are promising yet underexplored strategies. 
\subsection{\textit{Video playback}}
Recent research indicates that encouraging streaming users to adopt environmentally conscious preferences, such as accepting certain quality degradations to reduce \ch{CO2} emissions, can significantly impact the energy consumption of video streaming~\cite{hossfeld2023greener, bingol2023quality}. Therefore, the next improvement for CMCD and CMSD protocols or AI-based ABRs may involve incorporating energy-awareness information, enabling both video players and delivery components to provide green services. In addition, optimizing AI techniques for lower energy consumption on the client side, particularly for computationally intensive tasks such as video super-resolution, remains an open research topic to ensure efficiency and sustainability.
\subsection{\textit{VQA}}
AI-based VQA solutions are often perceived as black boxes, posing challenges in understanding the relationship between input factors and model outcomes~\cite{wehner2023explainable}. However, adopting Explainable AI (XAI) techniques~\cite{clement2024beyond,wehner2023explainable}can mitigate this challenge by providing clear explanations of the model's behavior and the rationale behind specific decisions. This is particularly crucial for energy-efficient video streaming, where transparent AI decisions can optimize energy consumption without compromising user experience.
\subsection{\textit{Generative AI}}
The emergence of Generative AI and Large Language Models (LLMs)~\cite{hoffmann2022training, chowdhery2023palm, touvron2023llama} is revolutionizing the video streaming lifecycle, similar to their impact on other applications~\cite{zhou2024survey}. Current integrations have reduced the complexity of conventional compression techniques~\cite{mentzer2022vct}, enabled players to adjust data transmission rates through more precise network bandwidth predictions~\cite{azmin2022bandwidth}, accelerated super-resolution models~\cite{mentzer2022vct}, and developed ABR algorithms customized for diverse network conditions~\cite{zhou2024survey}. However, the models behind Generative AIs and LLMs are massive and energy-intensive. However, the models behind Generative AIs and LLMs with numerous features are massive and energy-intensive. Recent research revealed that training a single LLM can emit up to \num{284} tons of \ch{CO2}, equivalent to the lifetime emissions of five cars, while inference processes also demand significant computational resources~\cite{strubell2019energy,bingol2023quality}. Consequently, a significant research gap exists concerning the energy efficiency of these methods for streaming use cases.

\section{Conclusion}
\label{sec:concl}
The increase in video consumption across various devices has raised concerns about the \ch{CO2} emissions attributed to video streaming, which require solutions to mitigate the environmental impact. This survey investigated AI-based solutions to reduce energy consumption in primary aspects of the video streaming lifecycle, \ie video encoding, delivery, and playback. It provided essential background on encoding, delivery networks, and playback as integral components of the video streaming lifecycle. It explored VQA and energy models and metrics. It also presented a taxonomy and a state-of-the-art review of \num{59} AI-based energy-aware solutions in four sections. Finally, it outlined potential future research directions, highlighting opportunities to improve energy efficiency in video streaming using AI methods.
\bibliographystyle{./bibliography/ACM-Reference-Format}
\bibliography{main.bib}
\printnomenclature
\end{document}